\documentclass{jfm}
\usepackage{graphicx}

\usepackage{orcidlink}
\usepackage{subcaption}
\usepackage{newtxtext}
\usepackage{newtxmath}
\usepackage{natbib}
\usepackage{hyperref}
\hypersetup{
    colorlinks = true,
    urlcolor   = blue,
    citecolor  = black,
}
\newcommand{\uad}{\boldsymbol{u}^{\dagger}}
\newcommand{\upe}{\boldsymbol{u}{'}}

\newcommand{\RomanNumeralCaps}[1]

\title{Nonlinear optimal perturbation growth in pulsatile pipe flow}

\author{P. Keuchel\aff{1}
\corresp{\email{patrick.keuchel@zarm.uni-bremen.de}},
\and M. Avila \aff{1}\aff{2}}
\affiliation{\aff{1} Center of Applied Space Technology and Microgravity (ZARM), University of Bremen, Am Fallturm 2, 28359 Bremen, Germany
\aff{2} MAPEX Center for Materials and Processes, University of Bremen,
Am Biologischen Garten 2, 28359 Bremen, Germany}

\begin{document}
\maketitle

\begin{abstract}
Pulsatile fluid flows through straight pipes undergo a sudden transition to turbulence that is extremely difficult to predict. The difficulty stems here from the linear Floquet stability of the laminar flow up to large Reynolds numbers, well above experimental observations of turbulent flow. This makes the instability problem fully nonlinear and thus dependent on the shape and amplitude of the flow perturbation, in addition to the Reynolds and Womersley numbers and the pulsation amplitude.
This problem can be tackled by optimizing over the space of all admissible perturbations to the laminar flow. In this paper, we present an adjoint optimization code, based on a GPU implementation of the pseudo--spectral Navier--Stokes solver \texttt{nspipe}, which incorporates an automatic, optimal check--pointing strategy. 
We leverage this code to show that the flow is susceptible to two distinct instability routes: One in the deceleration phase, where the flow is prone to oblique instabilities, and another during the acceleration phase with similar mechanisms as in steady pipe flow. Instability is energetically more likely in the deceleration phase. Specifically, localised oblique perturbations can optimally exploit nonlinear effects to gain over nine orders of magnitude in energy at a peak Reynolds number of $Re_{\max}\approx 4000$. These oblique perturbations saturate into regular flow patterns that decay in the acceleration phase or break down to turbulence depending on the flow parameters. In the acceleration phase, optimal perturbations are substantially less amplified, but generally trigger turbulence if their amplitude is sufficiently large.
\end{abstract}

\begin{keywords}
Authors should not enter keywords on the manuscript.
\end{keywords}

\section{Introduction}
Cardiovascular flows are pulsatile by nature and feature elastic vessels, non--Newtonian blood behaviour and complex geometries.
In the physiological context, the presence of turbulent or irregular flow patterns is particularly important, as regions of altering wall shear stresses are often linked to the onset of cardiovascular diseases \citep{Malek1999,Davies2009}.
Despite the relevance of turbulence in the cardiovascular system, however, the processes and conditions by which the flow transitions to turbulence are not understood in detail.
In order to isolate the influence of different features of the cardiovascular system, it is useful to study them individually. Specifically, we here asses the influence of finite--amplitude perturbations in pulsatile pipe flow and their role in the transition to turbulence. 
In addition to the Reynolds number $Re=D\bar{U}/\nu$, pulsatile flow in straight, smooth and rigid pipes is governed by the pulsation amplitude $A=U_{\max}/\bar{U}-1$ and the Womersley number $Wo=D/2\sqrt{2\pi f /\nu}$ \citep{Womersley1955}. Here $D$ is the pipe diameter, $\bar{U}$ the mean bulk velocity, $U_{\max}$ the maximum bulk velocity, $f$ the frequency of the pulsation and $\nu$ the kinematic viscosity.
\\
In the limiting case of steady pipe flow ($A=0$), the laminar flow is known to be linearly stable up to at least $Re>10^7$ \citep{Meseguer2003}. 
Pulsatile flows are linearly (Floquet) unstable already at lower $Re$ \citep{Thomas2011}, depending on the pulsation amplitude $A$ and frequency $Wo$. However, in experiments transition to turbulence is observed well before the linear stability threshold is crossed \citep{Trip2012, Xu2017, Xu2020, Brindise2024}.
The mechanisms by which perturbations can transiently extract energy from the laminar flow to grow have been widely studied in steady pipe flow and other shear flows \citep{Schmid2001}.
Generally, transient growth originates from the non--normality of the system and depends on the initial energy, $E_0$, and shape of the perturbation \citep{Trefethen1993,Schmid2001}. 
In the linear regime, $E_0\rightarrow 0$, \citet{Schmid1994} show that the optimal perturbation to steady pipe flow consists of a pair of streamwise vortices. These vortices generate streamwise streaks due to the lift--up effect \citep{Brandt2014}. Although these streaks are not able to cause turbulence by themselves \citep{Waleffe1995}, secondary instabilities can lead to streak breakdown and trigger transition \citep{Reddy1998,Meseguer2003_StreakBreakdown}. 
\\
The dynamics of pulsatile flows is much richer, because the laminar flow evolves in the viscous time scale, $D/\nu^2$, whereas the perturbation dynamics evolves in the convective time scales $D/\bar{U}$.
At low Womersley numbers, the laminar profile is parabolic and the pulsation period is long in terms of convective time units, $T=\pi Re /(2Wo^2)$. In this quasi--steady limit $T>>1$, optimal perturbations are as in steady pipe flow and their growth is mainly governed by the maximum Reynolds number.
At high Womersley numbers, perturbations cannot react to the fast pulsation and the steady case is also recovered, but with the growth controlled by the mean Reynolds number $Re$ \citep{Xu2017}.
For intermediate Womersley numbers $4 \lesssim Wo \lesssim 18$ and sufficiently large amplitudes $A \gtrsim 0.4$, the laminar profile features inflection points during the deceleration phase of the period and is instantaneously linearly unstable \citep{Moron2022}. The instantaneously most unstable mode has a helical shape and is attached to the unstable inflection point that locally fulfils the Fj\o rtoft criterion. After an initial energy boost via the Orr mechanism, the linear optimal perturbation leverages this modal mechanism and travels radially attached to the inflection point thereafter \citep{Moron2022}. Its energy growth scales exponentially with $Re$ \citep{Xu2021}, because the lifetime of the unstable inflection point scales linearly with $Re$ in convective time units \citep{Moron2022}
\\
These three distinct regimes of linear optimal perturbation dynamics according to the Womersley number, result in three corresponding nonlinear regimes. At low $Wo$ (quasi--steady case), transition to turbulence occurs when the instantaneous Reynolds number exceeds the natural transition number of the experiments in the steady case \citep{Stettler1986, Xu2018}. Subsequently, the flow stays turbulent, if the minimum Reynolds number is sufficiently large, or relaminarises resulting in a cyclic trigger--decay sequence. \citet{Xu2017} showed that this decay of turbulence is stochastic and a critical Reynolds number for sustained turbulence, which depends on $Wo$ and $A$, can only be defined in a statistical sense, exactly as in steady pipe flow \citep{Avila2011}. The high Womersley regime is again essentially as steady pipe flow and is controlled by the mean Reynolds number, because turbulence does not have time to react to the fast pulsation \citep{Trip2012,Xu2017,Xu2018}.
In the intermediate Womersley regime, a smooth transition between the two cases is observed at low $A\lesssim0.4$ \citep{Xu2017,Xu2018}.
\\
At high $A \gtrsim0.4$ the nonlinear dynamics is entirely different and \citet{Xu2020} observed that small geometric imperfections can trigger helical flow patterns during the deceleration phase that break down into turbulence before relaminarising during the acceleration phase. This process repeated itself cyclically in the experiments. Additionally they performed direct numerical simulations (DNS) of linearly optimal helical waves superimposed to the laminar flow at finite amplitude. The helical waves were found to be strongly amplified and, in the presence of background noise and a sufficient initial amplitude, to break down to turbulence. 
The effect of small geometric imperfections was further investigated numerically by \citet{Feldmann2020}, who perturbed the pulsatile flow with small axisymmetric, mirror symmetric and tilted bumps. Axisymmetric bumps were unable to trigger turbulence, whereas non--axisymmetric bumps triggered vigorous vortical structures during the deceleration phase.
Specifically, helical flow patterns were observed in the presence of the tilted bump and oblique patterns for the symmetric one.
\\
Although linear transient growth analyses capture the mechanism of perturbation amplification, it does not provide information on whether turbulence is triggered, nor which type of perturbation is most efficient in destabilizing the laminar flow. The answer to these questions requires a fully nonlinear approach, i.e. considering also the initial energy of the perturbation \citep{Kerswell2018}.
\\
Nonlinear optimal perturbations have been computed for different canonical systems, including steady pipe \citep{Pringle2010, Pringle2012, Pringle2015}, boundary--layer \citep{Cherubini2010, Cherubini2011} and plane Couette flows \citep{Monokrousos2011, Duguet2010, Duguet2013}. In all the systems, a qualitatively similar behaviour was observed. After passing a certain initial energy level, nonlinearities become substantial and modes can interact with each other. The corresponding optimal perturbations exploit nonlinearities and sequentially leverage different growth mechanisms to achieve a significantly larger growth than linear optimal perturbations. Specifically, they first leverage the Orr mechanism \citep{Orr1907} to experience a quick initial energy amplification. Subsequently, they evolve into oblique waves \citep{Schmid1992} and then into modes that use the lift--up mechanism \citep{Landahl1980,Brandt2014} to generate high and low speed streaks. For sufficiently low initial energies, the streaks remain stable and the perturbations decay. Above a critical energy $E_0^{\text{crit}}$, defining the minimal seed \citep{Kerswell2018}, the streaks become unstable and break down to turbulence.
\\
In this paper, we investigate the competition between helical perturbations leveraging the inflection points of the laminar profile and the classical lift--up perturbations leveraging the shear of the time--averaged profile. We focus on their role in destabilizing the laminar flow and eventually triggering turbulence.
For this purpose, we apply nonlinear non--modal stability theory \citep{Kerswell2018} and compute nonlinear optimal perturbations of pulsatile pipe flow.
In addition to the Reynolds number $Re$ and the initial energy of the perturbation $E_0$, optimal perturbations in pulsatile flow are governed by the pulsation amplitude $A$, Womersley number $Wo$ and initial perturbation time $\tau_0$. The latter is crucial because of the prominence of the lift--up effect in the acceleration phase and helical perturbations in the deceleration phase \citep{Xu2021}. 
\\
The rest of the paper is structured as follows.
The methodology, implementation and technical details are presented in section \S \ref{sec: Method and Implementation}.
In section \S \ref{sec: Results Helial vs Sym}, we quantify the role of linearly optimal helical and reflection symmetric (oblique) perturbations in the nonlinear regime.
We then perform nonlinear optimisations and study the effect of the initial amplitude on the shape and growth of optimal perturbations. In combination with direct numerical simulations, we quantify the role of different optimal perturbations in the transition to turbulence in pulsatile pipe flow. To do this end, in section \S \ref{sec: Results NLOPs} the parameters of experiments of \citet{Xu2020} are considered as reference and optimal perturbations at various initial energies and different times throughout the period are computed and analysed.
Finally, we explore the effect of different pulsation parameters $(Re, Wo, A)$ on transient amplification of optimal perturbations and the transition to turbulence at selected initial times.
In section \S \ref{sec: Discussion}, the results are discussed and compared to previous linear non--modal analyses, nonlinear direct numerical simulations and experimental results.

\section{Methods \& Implementation}
\label{sec: Method and Implementation}
We consider the flow of a fluid with constant density $\rho$ and kinematic viscosity $\nu$ through a cylindrical pipe  driven with a pulsatile harmonic mass flux. Experimentally, this is equivalent to a pulsatile flow driven by a piston. Throughout this paper, we scale all lengths with the pipes radius $D/2$ and velocities with twice the mean bulk velocity $2\bar{U}$. The steady component of the laminar profile then reads
\begin{equation}
    \boldsymbol{\bar{u}}(r)=(1-r^2)\boldsymbol{e}_z \hspace{10mm} r\in[0,1]
\end{equation}
and the enforced pulsatile bulk velocity is
\begin{equation}
   u_b(t)=\frac{1}{2}\left[1+A \sin\left(2\pi\frac{t}{T}\right)\right].
\end{equation}
The resulting temporally evolving laminar base flow profile $\boldsymbol{U}$ is given by the classical Sexl--Womersley solution \citep{Sexl1930, Womersley1955}.
We aim to find the global optimal perturbation $\boldsymbol{u}{'}(r,\theta,z,t=\tau_0)=\boldsymbol{u}_0{'}$ introduced at time $\tau_0$, which maximizes its energy growth
\begin{equation}
    G=\frac{\left\langle |\boldsymbol{u}'(t={\tau_0+\tau})|^2\right\rangle}{\left\langle|\boldsymbol{u}'_0|^2\right\rangle}
    =\frac{E(t={\tau_0+\tau})}{E_0}
    \label{eq: defintion energy gain}
\end{equation}
at a given time $\tau_0+\tau$. Here
\begin{equation}
    \langle...\rangle=\int_0^{L_z} \int_0^{2\pi} \int_0^1 \ (...)r \ dr \ d\theta \ dz
\end{equation} 
with pipe length $L_z$ and $E_0$ is fixed in each optimization, but changed across runs.
To maintain the desired Reynolds number, a zero bulk velocity of is enforced for the perturbation, i.e. $ \langle \upe(r,\theta,z,t)\cdot \boldsymbol{e}_z\rangle=0$ at all times. \\
We follow the nonlinear variational approach of \citet{Kerswell2014} and define a Lagrangian to enforce appropriate constraints. First, we constrain the optimisation to perturbations with initial energy $E_0$ through the Lagrangian multiplier $\lambda$. In addition, during their evolution, perturbations must fulfil momentum and mass conservation, and have zero bulk velocity. These constraints are applied through the Lagrangian multipliers $\boldsymbol{u}^{\dagger}=(u_r^{\dagger},u_{\theta}^{\dagger},u_z^{\dagger}), \ p^{\dagger}$ and $\Gamma$, respectively, known as the adjoint variables. The Lagrangian is accordingly defined as
\begin{eqnarray}
    \mathcal{L}&=&\left\langle\frac{1}{2} |\boldsymbol{u}'(t=\tau)|^2\right\rangle
    \nonumber \\
    && +\lambda \left[  \left\langle\frac{1}{2} |\boldsymbol{u}'_0|^2\right\rangle -E_0\right]
    \nonumber \\
    && +\int_{{\tau_0}}^{{\tau_0+\tau}} \left\langle  \boldsymbol{u}^{\dagger} \boldsymbol{\cdot} \left[ \frac{\partial \boldsymbol{u}'}{\partial t}+(\boldsymbol{U}\boldsymbol{\cdot} {\nabla})\boldsymbol{u}' +(\boldsymbol{u}'\boldsymbol{\cdot} {\nabla})\boldsymbol{U} +(\boldsymbol{u}'\boldsymbol{\cdot} {\nabla})\boldsymbol{u}' +\nabla p' -\frac{1}{Re}\nabla^2\boldsymbol{u}' \right]  \right\rangle \ dt
    \nonumber \\
    && +\int_{{\tau_0}}^{{\tau_0+\tau}} \left\langle  p^{\dagger} \left[\nabla \cdot \boldsymbol{u}' \right ]\right\rangle \ dt
    +\int_{{\tau_0}}^{{\tau_0+\tau}} \Gamma \left\langle  \boldsymbol{u}' \boldsymbol{\cdot} \boldsymbol{e}_z\right\rangle \ dt.
\end{eqnarray}
We treat the zero flux constraint as in \citet{Pringle2012} and split the pressure field into a spatially homogeneous time dependent pressure gradient $f_z(t)\boldsymbol{e}_z$, which maintains the zero bulk velocity of the perturbation and a spatially periodic perturbation pressure, so the total perturbation pressure is $p'=f_z(t)z\boldsymbol{e}_z+\hat{p}'$.
\\
For optimal perturbation fields, i.e. constrained extrema of the functional $\mathcal{L}$, the first variation of $\mathcal{L}$ must vanish. Taking the first variation with respect to $\boldsymbol{u}'$ and $p'$, integrating by parts and considering no--slip boundary conditions at the pipe wall $\boldsymbol{u}'(r=1,\theta,z,t)=0$, as well as periodicity in the axial and azimuthal directions, leads to the following set of equations \citep{Pringle2010, Cherubini2010, Monokrousos2011, Kerswell2014}:
\begin{itemize}
    \item initial energy constraint 
        \begin{eqnarray}
        \frac{\delta \mathcal{L}}{\delta \lambda}= \left[  \left\langle\frac{1}{2} |\boldsymbol{u}'_0|^2\right\rangle -E_0\right]  =0
        \label{eq: energy constraint}
        \end{eqnarray}
    \item direct Navier--Stokes equations
        \begin{equation}
        \frac{\delta \mathcal{L}}{\delta \boldsymbol{u}^{\dagger}}=  \frac{\partial \boldsymbol{u}'}{\partial t}+(\boldsymbol{U}\boldsymbol{\cdot} {\nabla})\boldsymbol{u}' +(\boldsymbol{u}'\boldsymbol{\cdot} {\nabla})\boldsymbol{U} +(\boldsymbol{u}'\boldsymbol{\cdot} {\nabla})\boldsymbol{u}' +\nabla \hat{p}'+f_z \boldsymbol{e}_z -\frac{1}{\Rey}\nabla^2\boldsymbol{u}'  =\boldsymbol{0}
        \label{eq: direct momentum}
        \end{equation}
        \begin{equation}
        \frac{\delta \mathcal{L}}{\delta p^{\dagger}}= \nabla \cdot \boldsymbol{u}' =0
        \label{eq: direct continuity}
        \end{equation}
    \item zero bulk velocity of the direct field
        \begin{equation}
        \frac{\delta \mathcal{L}}{\delta \Gamma}= \left\langle  \boldsymbol{u}' \boldsymbol{\cdot} \boldsymbol{e}_z\right\rangle   =0
        \label{eq: direct flux constraint}
        \end{equation}
    \item compatibility condition
        \begin{eqnarray}
        \frac{\delta \mathcal{L}}{\delta \boldsymbol{u}'(t={\tau_0+\tau})}= \boldsymbol{u}(t={\tau_0+\tau})+\boldsymbol{u}^{\dagger}(t={\tau_0+\tau})  = \boldsymbol{0}
        \label{eq: compatibility}
        \end{eqnarray}
    \item adjoint Navier--Stokes equations
        \begin{equation}
            \frac{\delta \mathcal{L}}{\delta \boldsymbol{u}'}=-\frac{\partial \boldsymbol{u}^{\dagger}}{\partial t}-\left[(\boldsymbol{U}+\boldsymbol{u}')\cdot \nabla\right]\uad+\left[\nabla (\boldsymbol{U}+\boldsymbol{u}')\right]^T\uad -\nabla \hat{p}^{\dagger}+f_z^{\dagger}\boldsymbol{e}_z-\frac{1}{Re}\nabla^2\uad =\boldsymbol{0}
            \label{eq: adjoint momentum}
        \end{equation}
        \begin{equation}
            \frac{\delta \mathcal{L}}{\delta \hat{p}'}=   \nabla \cdot \uad=0
            \label{eq: adjoint continuity}
        \end{equation}
    \item zero bulk velocity of the adjoint field
        \begin{equation}
            \frac{\delta \mathcal{L}}{\delta f_z}= \left\langle  \uad \boldsymbol{\cdot} \boldsymbol{e}_z\right\rangle     =0
            \label{eq: adjoint flux constraint}
        \end{equation}
    \item optimality condition
        \begin{eqnarray}
        \frac{\delta \mathcal{L}}{\delta \boldsymbol{u}'_0}= \lambda \upe_0 -\uad_0   =\boldsymbol{0}.
        \label{eq: optimality}
        \end{eqnarray}
\end{itemize}
Equations \eqref{eq: adjoint momentum}--\eqref{eq: adjoint continuity} are the adjoint Navier--Stokes equations, subject to a zero bulk adjoint velocity \eqref{eq: adjoint flux constraint}. Note the negative sign of the temporal derivative, which allows us to solve these equations backwards in time from $t=\tau_0+\tau$ to $t={\tau_0}$. The initial condition for the adjoint equations is given by the compatibility condition \eqref{eq: compatibility}, obtained by integrating the direct equations forward. As in the direct equations, the dual pressure field $p^{\dagger}$ is divided in a time dependent pressure gradient $f_z^{\dagger}\boldsymbol{e}_z$, which maintains the zero bulk velocity \eqref{eq: adjoint flux constraint} and a spatially periodic part $\hat{p}^{\dagger}$.
\subsection{Adjoint looping procedure}
As in previous studies (\citet{Pringle2010, Cherubini2010, Monokrousos2011, Duguet2013}), we solve the set of equations employing the iterative adjoint looping procedure. We initialise the first adjoint looping iteration $s=0$ with an initial guess of the optimal perturbation $\upe_0{^{(s=0)}}$ scaled to energy $E_0$. The Navier--Stokes equations \eqref{eq: direct momentum}--\eqref{eq: direct continuity} subjected to \eqref{eq: direct flux constraint} are then integrated forward in time, to obtain the velocity field $\boldsymbol{u}'(t=\tau_0+\tau)$. With the solution of the direct system at time $t={\tau_0+\tau}$, the adjoint field is initialized according to the compatibility condition \eqref{eq: compatibility}. The adjoint equations \eqref{eq: adjoint momentum}--\eqref{eq: adjoint continuity}, subjected to \eqref{eq: adjoint flux constraint}, are then solved backwards in time to obtain $\boldsymbol{u}^{\dagger}(t=0)$. 
Since the initial guess of the optimal perturbation $\boldsymbol{u}_0'{^{(s=0)}}$ is unlikely to be optimal and hence the optimality condition \eqref{eq: optimality} is not fulfilled,
\begin{equation}
    \nonumber
    \frac{\delta \mathcal{L}}{\delta \upe_0}\neq 0,
\end{equation}
the initial perturbation is improved as explained in the next section. 
With the improved initial perturbation $\boldsymbol{u}_0'{^{(s)}}$, a new iteration $s+1$ of the adjoint looping procedure is initialized and the procedure is repeated until the optimality condition is fulfilled.
\subsection{Optimization step}
There are several approaches to move the initial condition towards a maximum, such as gradient--ascent or rotational--gradient methods \citep{Foures2013, Kerswell2018}.
We employ a gradient--ascent method,
\begin{eqnarray}
\boldsymbol{u}_0'{^{(s+1)}}&&=\boldsymbol{u}_0'{^{(s)}}+\alpha \left[ \frac{\delta \mathcal{L}}{\delta \upe_0}\right]^{(s)}
    \\
    && =\boldsymbol{u}_0'{^{(s)}}+\alpha\left[\lambda \upe_0{^{(s)}} -\uad_0{^{(s)}}\right],
    \label{eq: perturbation update}
\end{eqnarray}
where $\alpha$ is the step size. The Lagrangian multiplier $\lambda$ is chosen to enforce the initial energy constraint of the updated perturbation
\begin{equation}
    E_0
    =\left\langle \frac{1}{2}|\upe_0{^{(s+1)}} |^2 \right\rangle
    =\left\langle \frac{1}{2}|(1+\lambda \alpha)\upe_0{^{(s)}}-\alpha \uad_0 {^{(s)}} |^2 \right\rangle.
\end{equation}
From this condition, we obtain the quadratic equation
\begin{equation}
    \lambda_\pm =\frac{\alpha C_0-2E_0}{2 \alpha E_0} \pm \sqrt{ \left( \frac{-\alpha C_0 +2E_0}{2 \alpha E_0}      \right)^2 -\frac{\alpha {{E}_0^{\dagger}}^{(s)} -C_0}{\alpha E_0}},
    \label{eq: lambda solution}
\end{equation}
where $E_0$ and ${E_0^{\dagger}}^{(s)}$ are the energies of the direct and adjoint field at $t=0$ respectively and $C_0=\left \langle  \upe_0{^{(s)}}\cdot \uad_0{^{(s)}}    \right \rangle$. 
Equation \eqref{eq: lambda solution} yields real solutions for step sizes provided that
\begin{equation}
    \alpha_{min}=-{\left(\frac{E_0^{\dagger}}{E_0}-\frac{C_0^2}{4E_0^2} \right)^{-1/2}} \leq \alpha \leq {\left(\frac{E_0^{\dagger}}{E_0}-\frac{C_0^2}{4E_0^2} \right)^{-1/2}}  = \alpha_{max}.
\end{equation}
As we seek maxima of $\mathcal{L}$, we must choose a positive step size of $0<\alpha \leq \alpha_{max}$. We here determine the step size $\alpha$ with a backtracking line search and choose $\lambda_+$, which yields updates close to the previous perturbation \citep{Foures2013}.
First, the initial perturbation is updated according to equation \eqref{eq: perturbation update} and \eqref{eq: lambda solution} and $\alpha=\alpha_{max}$. If the growth $G$ of the updated perturbation is higher, we proceed with the next adjoint iteration. If it yields a lower $G$, the step size is halved, a new update is computed and $G$ is evaluated again. This procedure is repeated until $G$ is increased or the relative change of the gain drop below a tolerance
\begin{equation}
    \left| \frac{G^{s+1}-G^{s}}{G^{s+1}} \right|  <10^{-8}.
\end{equation}

\subsection{Numerical solution of the direct and adjoint equations}
The direct and adjoint equations, are discretized using a Fourier--Galerkin method in the azimuthal and axial directions and high--order finite differences of order six in the radial direction. The $N_r$ radial points are initially located at the Chebyshev collocation points with a subsequent relaxation to locate more points away from the wall. All variables are expressed as
\begin{equation}
    f(r,\theta,z,t)= \Re \left[\sum_{k=-K}^{K} \sum_{m=-M}^{M} \hat{f}_{k,m}(r,t) e^{ik_0kz+im\theta}\right]
\end{equation}
with complex Fourier coefficients $ \hat{f}_{k,m}$, axial and azimuthal wave number $k$ and $m$ respectively and fundamental axial wave number $k_0=2\pi/L_z$.
The resolutions $(M,K)$ used for the simulations were adapted depending on the Reynolds number and length of the pipe so that the relative energy contained in the smallest axial and azimuthal modes was sufficiently small 
\begin{eqnarray}
        e_{M}=\frac{L_z}{2E_0}\sum_{k=-K}^{K} \int_0^1 |\hat{\upe}_{k,M}|^2 r \ dr<10^{-8} 
        \\
        e_K=\frac{L_{\theta}}{2E_0}\sum_{k=-M}^{M} \int_0^1 |\hat{\upe}_{K,m}|^2 r \ dr <10^{-8}.
\end{eqnarray}
A convergence study was performed to determine the necessary radial resolution at a constant Reynolds number of $Re=4000$, corresponding to the maximum instantaneous peak Reynolds number considered and the number of radial points was set to $N_r=64$ if not stated otherwise.
\\
In order to achieve the adjoint looping procedure, we extend the GPU version of the open source pseudo--spectral Navier--Stokes code \href{https://github.com/Mordered/nsPipe-GPU}{\texttt{nsPipe-GPU}} \citep{Moron2024}, based on the CPU \href{https://github.com/dfeldmann/nsCouette}{\texttt{nsPipe}} code \citep{Lopez2020}.
Time integration of the direct and adjoint system is performed using a second--order predictor corrector method with a fixed time step $\Delta t$. During the correction, the forcing term $f_z\boldsymbol{e}_z$ ($f_z^{\dagger}\boldsymbol{e}_z$) is adjusted to satisfy equation \eqref{eq: direct flux constraint}. The boundary conditions at the wall are imposed using an influence matrix method, following \texttt{openpipeflow} \citep{Willis2017}. For more details of the direct solution see \citet{Lopez2020} and \citet{Moron2024}.

\subsection{Optimal check pointing schedule}
In order to compute the convective term in the backward integration of the adjoint equations \eqref{eq: adjoint momentum}, the solution of the direct problem $\upe$ is required at every time step. Hence, from a performance point of view, it is desirable to store all direct fields in the GPU memory. However, due to the limited memory of the GPU, this is generally infeasible and the direct field must be recomputed from checkpoints. Let $N_G$ denote the maximum number of checkpoints that can be stored in memory. Then, the question arises of when velocity fields should be stored in the memory as checkpoints during the forward integration. \citet{Griewank2010} showed that a logarithmic spacing of checkpoints minimizes the number of forward time steps necessary to recover all direct fields during the backward integration of the adjoint system. 
Thereby, checkpoints that are no longer needed, are overwritten and updated during the intermediate forward integrations.
Following this procedure, the minimum number of intermediate steps $N_{\text{rec}}$ to recover all forward fields is
\begin{equation}
\label{Eq: total recovery steps}
N_{\text{rec}}=SN_{\text{steps}}-\beta(N_{G}+1,S-1),
\end{equation}
where $N_{\text{steps}}$ is the total number of forward time steps and $S$ a unique integer, for which
\begin{equation}
\beta(N_{G},S-1) < N_{\text{steps}} \leq \beta(N_{G},S), 
\label{eq: CP S condition}
\end{equation}
with
\begin{equation}
\beta(N_{G},S)=\left( \begin{array}{c}
N_{G}+S\\
N_{G}
\end{array}\right)
=\frac{(N_{G}+S)(N_{G}+S-1)\dots(S+1)}{N_{G}!}.
\end{equation}
For large scale and memory demanding simulations only a small number of memory checkpoints can be stored. Hence, it may become advantageous to additionally store checkpoints to disc.
We developed a check pointing procedure, which combines the optimal logarithmic spacing of memory checkpoints \citep{Griewank2010} with additional checkpoints saved to disc.
In this procedure, we first save $N_F$ equispaced fields to disc during the forward integration.
Additionally, in the last interval, i.e between the last checkpoint saved to disc and the end of the forward integration, $N_G$ memory checkpoints are saved with a logarithmic spacing.
Then, during the backward integration of the adjoint equations, the logarithmically spaced memory checkpoints are updated as in \citet{Griewank2010}. 
This procedure is repeated for each intermediate integration between disc checkpoints, until $t=0$ is reached.
We find the optimal number of file checkpoints, $N_F$, by minimizing the total computing time to recover all direct fields
\begin{align}
\label{Eq: total recovery time}
\min\left( T^{\text{rec}} \right)=N_{F}T^{\text{i/o}}+\left \{ (N_F+1)N_{\text{rec}}-N_{\text{init}} \right \} T^{\Delta t},
\end{align}
where $T^{\text{i/o}}$ is the time it takes to read/write a field from/to file and $T^{\Delta t}$ the time it takes to perform a single time step of the forward integration. $N_{\text{init}}$ is the number of time steps between the last file checkpoint and last memory checkpoint and takes into account that the initial set of memory checkpoints is already saved during the initial forward integration.
The times $T^{\text{i/o}}$ and $T^{\Delta t}$ depend on the hardware and for a logarithmic spacing the optimal $N_{\text{rec}}$ is given by equation \eqref{Eq: total recovery steps}.
Assuming that the space on disc and thus the number of file checkpoints is not limited, a recycling of file checkpoints, as employed for the memory checkpoints, does not provide any advantage and therefore an equidistant spacing is optimal.
As we compute the number of file checkpoints that minimises the recovery time, this strategy maximises the efficiency, i.e. it automatically minimises the run time for a given computer architecture.

\subsection{Verifications and parameter choices}
\label{sec: technicalities}
In order to validate our code, we reproduced the computations of nonlinear optimal perturbations to laminar steady pipe flow \citep{Pringle2012} and linear optimal perturbation in pulsatile pipe flow \citep{Xu2021}. Qualitative and quantitative comparisons and the convergence behaviour of our implementation are detailed in appendix \ref{app: Validation pulsatile}. 
\\
Throughout this paper, the pipe length is set to $L_z=50$, to delay the self interaction of axially localised perturbations due to the periodic boundary conditions. We found that $L_z=50$ is sufficiently long so that the maximum energy amplification of nonlinear optimal perturbations in steady pipe flow is independent of the pipe length, in agreement with \citet{Pringle2015}.
Given the large number of parameters governing the problem $(Re,Wo,A,\tau_0,E_0)$ and the computational cost of nonlinear optimisations, we compute optimal perturbations for selected parameter combinations and various $E_0$.
Based on linear transient growth analysis \citep{Xu2021}, linear optimal perturbations are either classical or helical perturbations, depending on $(Wo,A,\tau_0)$. Here we consider representative cases from the three characteristic regimes of pulsatile pipe flow and analyse the influence the initial perturbation energy and nonlinear effects.
\\
In addition to the pulsation and perturbation parameters $(Re,Wo,A,\tau_0,E_0)$, one must also specify the optimization time $\tau$, see eq. \eqref{eq: defintion energy gain}.
In the linear regime, i.e. small $E_0$, the time of maximum amplification $\tau^{(lin.)}$ is well known \citep{Xu2021,Pier2021} and we initially set the optimisation time to $\tau=\tau^{(lin.)}$. 
In the nonlinear regime we perform optimisations at selected $\tau$ and observe that the maximum energy amplification is usually reached near $\tau^{(lin.)}$. 
The structures of computed optimal perturbations are qualitatively robust in a certain range of $\tau$ and show only small deviations in their maximum amplification.
If $\tau$ is too short, optimal perturbations are characterised by the Orr mechanism and multiple streamwise vortices that quickly lift up fluid from the near wall regions to form high and low-speed streaks. These perturbations grow rapidly, but reach a smaller amplification due to their relatively large wave numbers.
For too long optimization times, optimal perturbations decay slowly and are characterised by low wave numbers.
\\
To verify the robustness of our results, we initialised selected optimisations with different initial guesses: (i) a rescaled turbulent field from a DNS of steady pipe flow at $Re=3000$, (ii) a helical perturbation with prescribed azimuthal velocity
\begin{equation}
    u_\theta(r,\theta)=(1-r^2)r^4\cos(4\pi r + \theta)
\end{equation}
(continuity was enforced in the first time step)
and (iii) a nonlinear optimal of steady pipe flow at $Re=2400$. For sufficiently small energies $E_0$, the different initial conditions lead to the same well--known linear optimal perturbations and for sufficiently large initial energies, to identical nonlinear optimal perturbations.
However, for initial energies, for which the nonlinear and linear optimal perturbations experience a similar growth, the results can depend on the initial guess and we either obtained the linear or nonlinear optimal perturbation.

\section{Linear optima in the nonlinear regime} 
\label{sec: Results Helial vs Sym}
\begin{figure}
    \centering
    \begin{subfigure}[b]{0.99\textwidth}
       \includegraphics[width=0.99\linewidth]{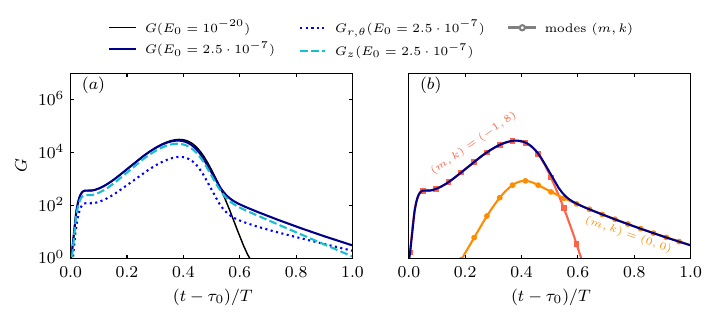}
     \end{subfigure}
    \begin{subfigure}[b]{0.99\textwidth}
       \includegraphics[width=0.99\linewidth]{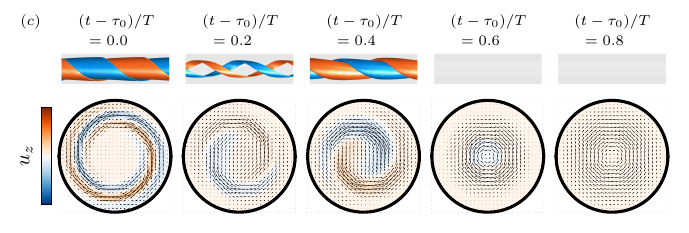}
     \end{subfigure}
    \caption{$(a)$ Growth $G$ (---$\!$---) of the linear optimal helical perturbation rescaled to $E_0=2.5\cdot10^{-7}$, further decomposed into the energy growth in the axial $G_{z}$ ($---$) and cross--section components $G_{r,\theta}$ ($\cdots \cdots$), together with the linear optimal ($E_0=10^{-20}$). $(b)$ Growth of selected Fourier modes of the rescaled helical wave with the overall growth as a reference. $(c)$ Isocontours of $u_z=\pm 6\cdot 10^{-3}$ and colormaps of the axial velocity in the range $u_z\in(-10^{-1},10^{-1})$, with quivers of the cross--sectional velocities $(u_r,u_\theta)$ at selected times throughout the period. Due to the significantly lower initial energy, at $t=0$, the isocontour limits are $u_z=\pm 6\cdot 10^{-5}$ and the contour limits are  $u_z\in(-2.5 \cdot 10^{-3},2.5 \cdot 10^{-3})$. The flow is from left to right and the resolution is set to $(N_r,M,K)=(64,40,181)$.
    }
    \label{fig: rescaled helical wave}
\end{figure}

\begin{figure}
    \centering
    \begin{subfigure}[b]{0.99\textwidth}
       \includegraphics[width=0.99\linewidth]{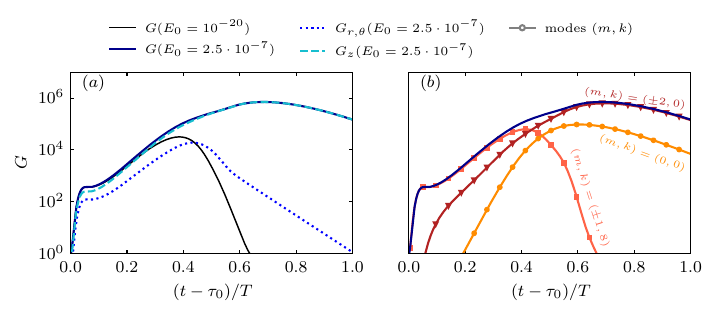}
     \end{subfigure}
    \begin{subfigure}[b]{0.99\textwidth}
       \includegraphics[width=0.99\linewidth]{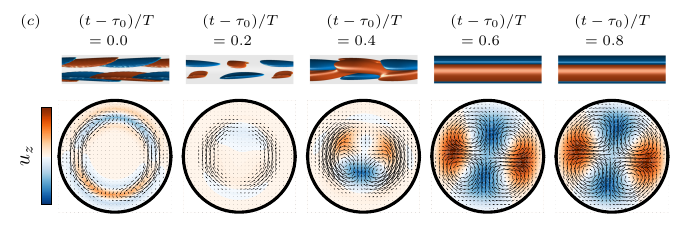}
     \end{subfigure}
    \caption{Same as in figure \ref{fig: rescaled helical wave} for an oblique perturbation consisting of the superposition of optimal helical perturbations $(m,k)=(\pm1,8)$.
    }
    \label{fig: rescaled standing wave}
\end{figure}

The linear transient growth analysis of \citet{Xu2021} showed that certain helical perturbations with azimuthal wave number $m=\pm1$ are optimal in the deceleration phase.
In the linear regime, Fourier modes evolve independently of each other, so any linear combination of two identical perturbations with opposite $m$ reaches the same amplification. However, at finite perturbation energies, Fourier modes interact nonlinearly and accordingly the growth may depend on the relative energetic contribution of the two helical components, $m=\pm1$.
We first considered the nonlinear evolution of linear optimal perturbations introduced during the deceleration phase ($\tau_0=T/2$) and scaled to finite energies.
We set the parameters to $(Re,Wo,A)=(2200,5.6,0.85)$, as in the experiments of \citet{Xu2020}.
At low initial energies ($E_0=10^{-20}$), we found that any linear combination of two optimal helical perturbations with modes $(m,k)=(\pm1,8)$ follows the same evolution (compare the black lines in figures \ref{fig: rescaled helical wave}$a$ and \ref{fig: rescaled standing wave}$a$) as expected in the linear regime.
\\
The blue solid line in figure \ref{fig: rescaled helical wave} depicts the temporal evolution of the energy growth of the optimal helical perturbation with $(m,k)=(-1,8)$ scaled to $E_0=2.5\cdot10^{-7}$.
For $t-\tau_0\lesssim0.5$, the energy growth is nearly identical to the linear case, i.e. the perturbation experiences a strong initial boost via the Orr mechanism, followed by an exponential growth due to the instantaneous linear instability of the laminar profile \citep{Moron2022}.
As seen in figure figure \ref{fig: rescaled helical wave}$a$, approximately the same maximum growth ($G_{\max}\approx 3.16\cdot10^4$ at $t-\tau_0=0.38T$) as in the linear regime is achieved.
As shown in figure \ref{fig: rescaled helical wave}$b$, the growth of the $(m,k)=(-1,8)$ mode abruptly ends in the acceleration phase, when the inflection point of the laminar profile disappears.
However, it nonlinearly interacts with itself and in secondary interactions  transfers energy into the $(m,k)=(0,0)$ mode thereby modifying the spatially averaged flow profile.
This helical perturbation can not leverage nonlinearities to boost its energy growth because its helical cross--flow vortices are aligned with the helical axial high-- and low--speed layers (see snapshots in figure \ref{fig: rescaled helical wave}). As a consequence, the lift--up effect is not efficient.
\\
In figure \ref{fig: rescaled standing wave}, we show the evolution of a reflection symmetric pair of oblique perturbations with an equal contribution of the linear optimal helical modes $(m,k)=(\pm1,8)$.
Similarly to the helical perturbation, the oblique perturbation first leverages the Orr mechanism and then grows exponentially due to the instantaneous linear instability of the base flow \citep{Moron2022}.
During the acceleration phase (after $t-\tau_0\approx 0.38T$), the modes $(m,k)=(\pm1,8)$ rapidly decay, however, by nonlinear interaction in their exponential growth phase they efficiently transfer energy into the modes $(m,k)=(\pm2,0)$, which then continue to grow in the acceleration phase. 
In the flow fields in figure \ref{fig: rescaled standing wave}$c$, this energy transfer is reflected by the radial convection of the axial velocity layers. In contrast to the linear case, radial velocities are now sufficiently large to convect these layers in the cross--section. Specifically, the initial pair of axial velocity layers is split by two pairs of counterrotating cross--flow vortices, which convect high momentum fluid from the centre of the nearly parabolic base flow to the near wall regions and low momentum fluid from the near wall base flow to the central regions. 
After $t-\tau_0\approx T/2$, solely the axial velocity component grows at the expense of the cross--sectional components (see figure \ref{fig: rescaled standing wave}$a$). 
This is characteristic of the lift--up effect, i.e. the perturbation extracts energy from the base flow via the term $u_r( \partial U_z/ \partial r)$ and further grows during the acceleration phase. 
The streaks remain stable and after the lift--up mechanism comes to an end at about  $t-\tau_0\approx 0.685T$, resulting in $G_{\max }\approx 7.08\cdot 10^{5}$, the perturbation decays viscously and transition does not occur.
\\
In summary, rescaled linear optimal perturbations initially grow by the same mechanisms as in the linear regime, i.e. the Orr mechanism followed by an exponential growth due to the instantaneous linear instability of the pulsatile base flow \citep{Xu2021,Moron2022}. While helical perturbations can not exploit nonlinearities to experience a higher energy amplification, the strong nonlinear interaction of a symmetric pair of helical perturbations transfers a sufficient amount of energy into the $(\pm2,0)$ modes, which allows for a further growth during the acceleration phase via the lift--up effect.
This oblique wave mechanism is well--known to produce a large transient \citep{Schmid1992} and is a hallmark of nonlinear optimal perturbations in other shear flows \citep{Cherubini2011,Pringle2012, Duguet2013}.

\section{Nonlinear optimal perturbation in pulsatile pipe flow}
\label{sec: Results NLOPs}
In this section, we present the results of our nonlinear non--modal stability analysis. In \S \ \ref{sec: NLOP Example}, we compute the nonlinear optimal perturbation for the same configuration as in the proceeding section. Subsequently, we study the effect of the initial energy (\S \ \ref{sec: effect of initial energy}), the perturbation time (\S \ \ref{sec: Effect of initial time}) and the flow parameters (\S \ \ref{sec: Paramterstudy}). Finally, we discuss the spatio-temporal dynamics of pulsatile turbulence triggered by minimal seeds at different pulsation parameters (\S \ \ref{sec: Turbulent dynamics}).

\subsection{Nonlinear optimal perturbation in the deceleration phase}
\label{sec: NLOP Example}
We introduced perturbations with initial energy $E_0=2.5 \cdot 10^{-7}$ at time $\tau_0=T/2$ and optimised their energy growth at the final time $t_f=\tau_0+\tau=T$. 
\begin{figure}
    \centering
    \begin{subfigure}[b]{0.99\textwidth}
       \includegraphics[width=0.99\linewidth]{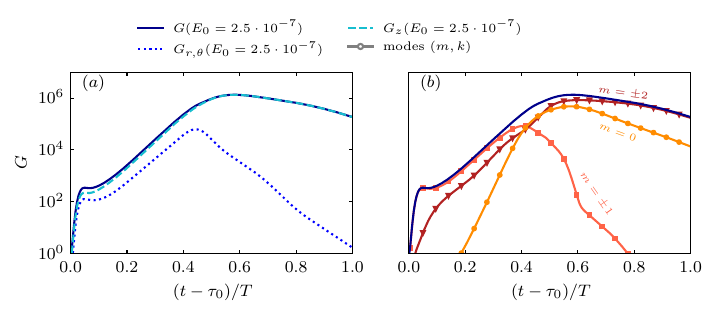}
     \end{subfigure}
    \begin{subfigure}[b]{0.99\textwidth}
       \includegraphics[width=0.99\linewidth]{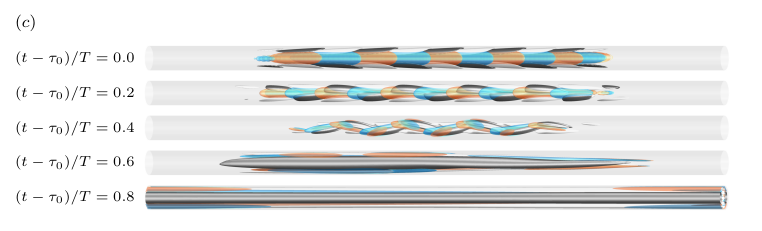}
     \end{subfigure}
    \caption{$(a)$ Growth $G$ (---$\!$---) of the nonlinear optimal perturbation for $(Re,Wo,A,\tau_0,\tau)=(2200,5.6,0.85,T/2,T/2)$ with initial energy $E_0=2.5\cdot10^{-7}$ , and its axial $G_{z}$ ($---$) and cross--section components $G_{r,\theta}$ ($\cdots \cdots$). $(b)$ Growth of selected modes with the total growth as a reference. For the azimuthal modes $m=\pm1$ we consider the axial modes $k\in(7,8,9)$ and for $m=0$ and $m=\pm2$ we consider $k\in(0,1,2)$ which contain most of the perturbation energy.
    $(c)$ Isocontours of $u_z=\pm 0.4u_z^{\max}$ (black and white) and $\omega_z=\pm0.4 \omega_z^{\max}$ (blue and orange). The flow is from left to right and the resolution was set to $(N_r,M,K)=(64,40,181)$.
    }
    \label{fig: NLOP deceleration}
\end{figure}
In figure \ref{fig: NLOP deceleration}$a$, we show the energy growth of the nonlinear optimal perturbation (NLOP) together with selected snapshots of its axial velocity field \ref{fig: NLOP deceleration}$b$. 
For our three initial guesses (see \S \ \ref{sec: technicalities}), the optimisation procedure converged to the same symmetric oblique perturbation with an equal contribution of the azimuthal modes $m=\pm1$ and localized along the pipe (see figure \ref{fig: NLOP deceleration}$c$).
In the spectra, this is characterised by the distribution of energy into several axial modes. Specifically, $95\%$ of the energy is initially contained in the axial modes $k\in(6,10)$.
Since the energy is now distributed in a large number of axial modes that interact with each other, nonlinearities transfer energy into many more modes. However, most of the energy is transferred into the modes $m=\pm2$ with $k\in(0,4)$  (see figure \ref{fig: NLOP deceleration}$b$).
Qualitatively, this NLOP grows by the same mechanisms as the linearly optimal symmetric oblique perturbation, shown in figure \ref{fig: rescaled standing wave}:
$(1)$ Initial energy boost by the Orr mechanism until $t-\tau_0\approx 0.05$ $(2)$ Exponential growth of various linearly unstable modes with azimuthal wave numbers $m=\pm1$ (see figure \ref{fig: NLOP deceleration}$b$) until about $t-\tau_0\approx 0.4$ $(3)$ Simultaneous nonlinear interaction of symmetric pairs of oblique modes transferring energy into $m=\pm2$ of lower axial wavenumber $(4)$ streak formation due to the lift--up effect.
Due to the axial localisation, the amplitude of the NLOP is locally much higher and results in stronger cross--flow vortices (compare the dotted lines in figure  \ref{fig: rescaled standing wave}$a$ and \ref{fig: NLOP deceleration}$a$). This enhances and extends the lift--up phase and ultimately leads to higher growth of the NLOP ($G^{\text{max}}\approx1.34\cdot10^6$ compared to $G^{\text{max}}\approx0.708\cdot10^6$ for the rescaled symmetric linear optimal). 
In figure \ref{fig: NLOP deceleration}$b$ the stronger effect of nonlinearities is reflected in an earlier and increased growth of the azimuthal modes $m=(0,\pm2)$.
\\
This behaviour is qualitatively similar to NLOPs in other shear flows, which leverage the same combination of the Orr mechanism, oblique wave interaction and lift--up mechanism, while localising in space in order to enhance the effect of nonlinearities \citep{Cherubini2011,Pringle2012, Duguet2013}. A noteworthy difference, however, is that the NLOP in pulsatile flow exploits the inflection points of the flow profile to significantly exceed the growth of optimal perturbations in steady flows.

\subsection{Effect of the initial energy}
\label{sec: effect of initial energy}
\begin{figure}
    \centering
    \includegraphics[width=0.99\linewidth]{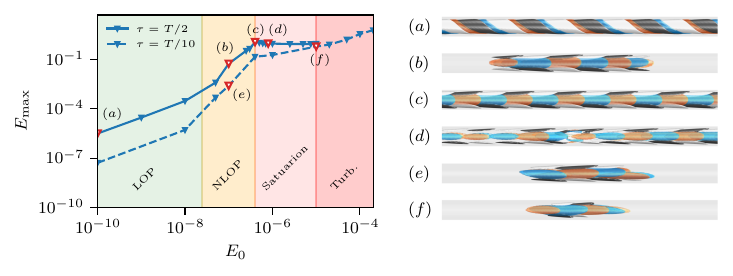}
    \caption{The achieved maximum absolute energy $E_{\max}$ of the optimal perturbation at different initial energies $E_0$. Solid and dashed lines depict the optimal energy growth for an optimisation time of $\tau=T/2$ and $\tau=T/10$ respectively. Shaded areas indicate the regime of linear optimal perturbations (LOP), localised symmetric oblique perturbations (NLOP), the non--turbulent saturation regime (Saturation) and the turbulent regime (Turb.). $(a)-(f)$ depict characteristic optimal perturbations in the different regimes.
    }
    \label{fig: deceleration summary}
\end{figure}

\begin{figure}
    \centering
    \includegraphics[width=0.99\linewidth]{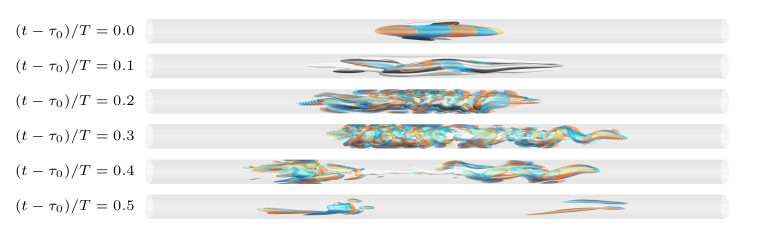}
    \caption{Evolution of the nonlinear optimal perturbation that first triggers transition at $(Re,Wo,A,\tau_0)=(2200,5.6,0.85,T/2)$. Isocontours of $u_z=(\pm0.0015,\pm0.015)$ (black and white) and $\omega_z=(0.015,\pm 0.15)$ (blue and orange) for $t/T=(0.0,0.1)$, respectively. For $t/T>0.1$ isocontour levels are $u_z=\pm0.15$ (black and white) and $\omega_z=\pm 0.15$ (blue and orange). The flow is from left to right and the resolution was set to $(N_r,M,K)=(64,40,181)$.}
    \label{fig: puff decay}
\end{figure}

We performed additional optimisations at various initial energies $E_0$ and summarise the results in figure \ref{fig: deceleration summary}.
For sufficiently low initial energies, we obtained linear optimal perturbations (LOPs) consisting of linear combinations of helical perturbations with $(m,k)=(\pm1,8)$ and for which the maximum energy scales linearly with the initial energy $E_{\max} \propto E_0$.
With increasing $E_0$, similar axially localised reflection symmetric oblique perturbation were found to be optimal as discussed in section \S \ref{sec: NLOP Example}. They further localise axially to locally enhance nonlinearities and significantly outgrow LOPs with an extended lift--up phase.
For $E_0 \gtrsim 2.5\cdot 10^{-7}$, the streak amplitude saturates due to a combination of strong nonlinearities and a very low minimum Reynolds number of $Re_{\min}=330$ at the end of the deceleration phase.
First, the nonlinear term shifts energy into large wave number modes, which rapidly dissipate it due to the low instantaneous Reynolds number. 
Second, the spatially averaged velocity profile is flattened, as nonlinearities transfer energy into the $(m,k)=(0,0)$ and the production $P=u_ru_z (\partial U_z/\partial r)$ is diminished as a result. When the instantaneous Reynolds number is sufficiently large again, the flow is almost streamwise independent and the residual fluctuations do not suffice to destabilize the flow.
Due to nonlinear saturation of the streak amplitude, adding more energy locally does not increase the maximum achieved energy. This introduces an opposed trend regarding the localisation: The optimisation procedure now starts to distribute the energy along the pipe without locally exceeding the amplitude threshold leading to streak saturation. 
Finally, at $E_0\approx4\cdot 10^{-7}$, the maximum growth is achieved by symmetric oblique perturbation filling the entire pipe (see figure \ref{fig: deceleration summary}$c$).
Further increasing the energy does no longer lead to larger growth as the saturation limit is reached in the entire pipe and no turbulence is triggered. In this regime, the optimisation procedure failed, as the objective function is locally flat and various initial conditions lead to the same maximum energy. The optimisation procedure returned perturbations that reach the saturation energy at $t=\tau_0+\tau$ and consist of a combination of the initial guess and the NLOP. In figure \ref{fig: deceleration summary}$d$, we show the result of an optimisation initialised with the NLOP of steady pipe flow. This perturbation features characteristics of the oblique pattern superposed to the structure of the initial guess.
\\
The only way that a perturbation introduced at $\tau_0=T/2$ can cause turbulence at these parameters is to grow extremely fast and trigger strong nonlinearities before viscous effects quickly act during the low Reynolds number phase.
We show this by performing additional computation with a short optimisation time of $\tau=T/10$.
Here, NLOPs further localise in axial and azimuthal direction and are characterised by four cross--sectional vortices (compare figures \ref{fig: deceleration summary}$b$ and $e$). They combine the Orr and lift--up mechanisms from the beginning to grow much faster. 
At small to moderate $E_0$, their maximum growth is smaller than optimal for $\tau=T/2$ (see figure \ref{fig: deceleration summary}).
However, at sufficiently large $E_0$ (see figure \ref{fig: deceleration summary}$f$), they are able to trigger a turbulent puff.
This puff quickly decays during the low Reynolds number phase, in agreement with the experimental and numerical observations by \citet{Xu2017, Xu2018}. The dynamics of transition and decay are illustrated in figure \ref{fig: puff decay}.
\\
In summary, we find that optimal perturbations in the deceleration phase experience a enormous energy growth and lead to strong symmetric flow patterns. Triggering turbulence, however, requires a significantly higher energy ($E_0\gtrsim 5\cdot 10^{-5}$ for the minimal seed at this parameters) due to the small instantaneous Reynolds number. 

\subsection{Effect of the initial perturbation time}
\label{sec: Effect of initial time}
In this section, we analyse the effect of different initial perturbation times $\tau_0$ on the optimal perturbations. 
For this purpose, we introduced a perturbation at the middle, end and beginning of the acceleration phase, $\tau_0=0, \ T/4$ and $3T/4$, respectively and computed the optimal perturbation with maximum gain after $\tau=T/2$ at various initial energies $E_0$.
\\
\begin{figure}
    \centering
    \includegraphics[width=0.99\linewidth]{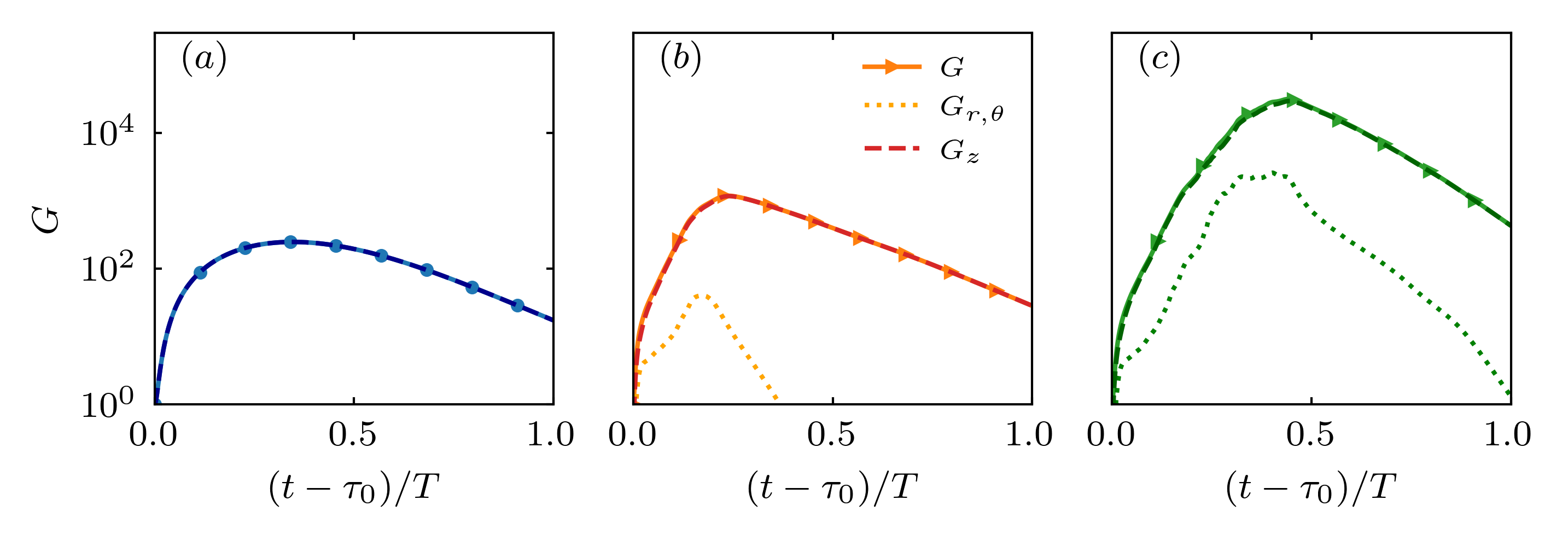}
    \caption{Growth $G$ (---$\!$---) of the optimal perturbation with initial energy $(a)$$E_0=1\cdot10^{-10}$ (linear optimum), $(b)$ $E_0=1.5\cdot10^{-4}$ (nonlinear optimum) and $(c)$ $E_0=2\cdot10^{-4}$ (above the minimal seed), at $(Re,Wo,A)=(2200,5.6,0.85)$ introduced at $\tau_0=0$. The total growth rate is further decomposed into the axial component $G_z$ ($---$) and the cross--sectional component $G_{r,\theta}$ ($\cdots \cdots$).}
    \label{fig: Pulsatile Accel Grwoth}
\end{figure}
For $\tau_0=0$ and sufficiently small initial energies, the optimal perturbation is the classical pair of streamwise vortices, known from steady pipe flow and in line with the results of \citet{Xu2021}.
These streamwise vortices induce a pair of axial high-- and low--speed streak that reach a maximum growth of $G_{\max}=247$ at $t^*=0.346T$ (see figure \ref{fig: Pulsatile Accel Grwoth}$a$). 
A nonlinear optimal first emerges at about $E_0=1.5\cdot10^{-4}$, which is three orders of magnitude higher than in the deceleration phase. This nonlinear optimal reaches a significantly higher maximum growth, $G_{\max}=1183$ at $t^*=0.238T$ (see figure \ref{fig: Pulsatile Accel Grwoth}$b$). As shown in figure \ref{fig: AC NLOP Snapshots E1.5e-4}$a$, this nonlinear optimum localizes in all directions. It is characterised by high-- and low--speed axial velocity layers, tilted against the shear of the base flow as well as two localised counter rotating vortices close to the wall. 
Qualitatively, this NLOP is as in steady pipe flow.
\begin{figure}
    \centering
    \includegraphics[width=0.99\linewidth]{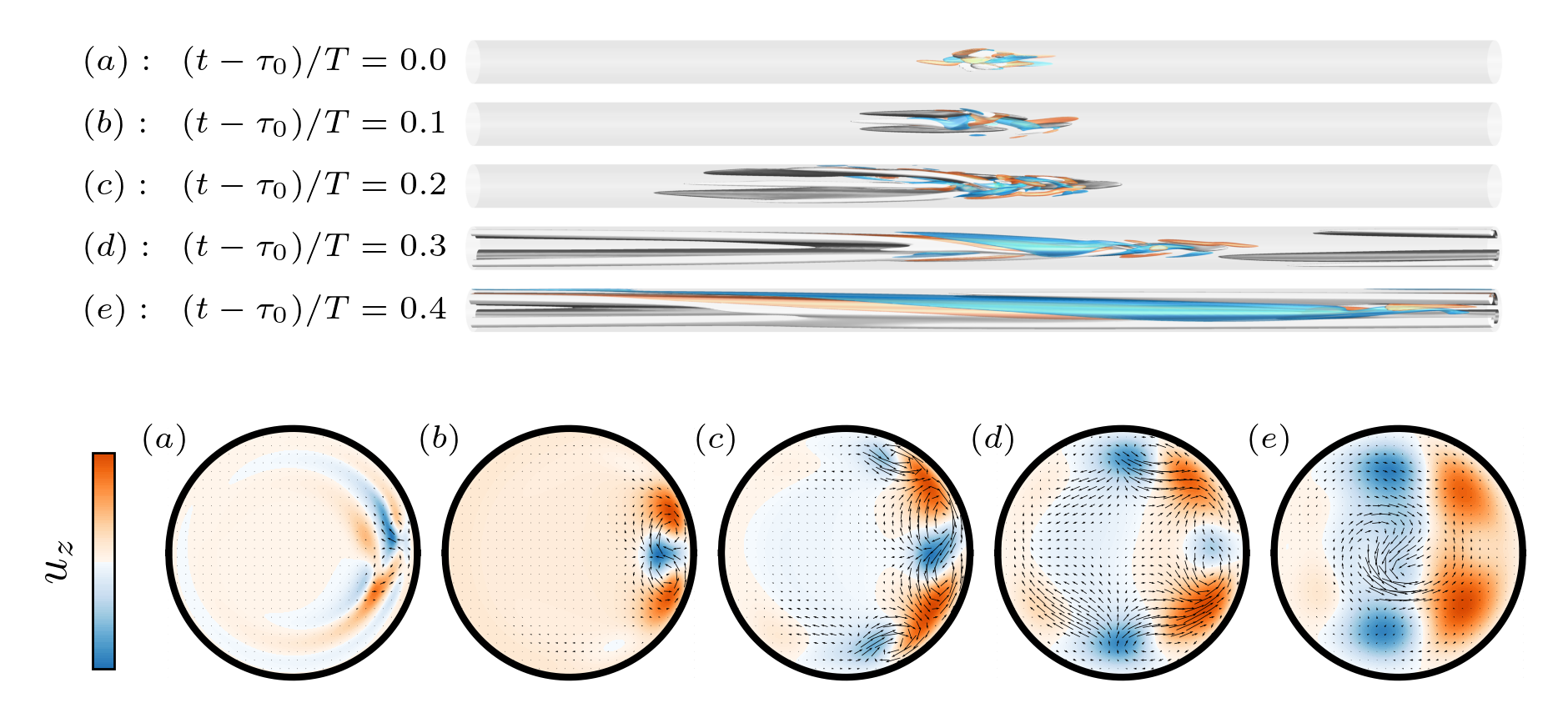}
    \caption{Nonlinear optimal perturbation with initial energy $E_0=1.5\cdot10^{-4}$, introduced at $\tau_0=0$ and optimised to experience a maximum growth at $t-\tau_0=T/2$. Isocontours indicate $u_z=\pm 0.4u_z^{\max}$ (black and white) and $\omega_z=\pm0.2 \omega_z^{\max}$ (blue and orange) and the corresponding contours of the axial velocity together with quivers of the cross--sectional flow are located at the axial position
    of maximum cross--sectionally integrated energy. The flow is from left to right and the resolution was set to $(N_r,M,K)=(64,40,181)$.
    }
    \label{fig: AC NLOP Snapshots E1.5e-4}
\end{figure}
As in the deceleration phase, the perturbation initially utilises the Orr mechanism to quickly gain energy in the axial and cross--sectional components. Once the axial high-- and low--speed layers align with the base flow shear, the cross--sectional vortices induce high-- and low--speed streaks due to the lift--up effect (see figure \ref{fig: AC NLOP Snapshots E1.5e-4}$b$). These vortices are localised close the wall, aligning with the large shear of the instantaneous laminar profile. During its growth phase, the perturbation unwraps and forms additional vortices close to the wall. The additional vortices induce an additional pair of low--speed streaks, as show in figure \ref{fig: AC NLOP Snapshots E1.5e-4}$c,d$, that quickly fill the entire domain. As shown in figure \ref{fig: Pulsatile Accel Grwoth}$b$, then the cross--sectional energy starts to decrease. For a short time, the axial component of the perturbation keeps growing due to the lift--up effect, before exponentially decaying in the deceleration phase.
Even though the shape of the NLOPs in the acceleration phase is very different from NLOPs in the deceleration phase, they leverage the same linear growth mechanisms (Orr and lift--up mechanism) coupled by nonlinear interaction of oblique modes. The essential difference is that the profile is instantaneously linear stable during the acceleration phase, which leads to a significantly lower energy growth.
\\
For $E_0\geq 2\cdot 10^{-4}$ (see figure \ref{fig: Pulsatile Accel Grwoth}$c$), NLOPs in the acceleration trigger transition to turbulence.
The localised streaks become unstable and break down to a turbulent puff shortly after passing the maximum instantaneous Reynolds number. This turbulent puff quickly elongates and fills the entire domain, before collapsing in the deceleration phase. The dynamics is very similar as reported by \citet{Xu2018} (see their figure 4). For the parameter combination considered here, the minimum Reynolds number is too low. Viscous effects dominate during the deceleration phase, turbulence decays and is not retriggered in the following period.
\\
In figure \ref{fig: Grwoth rate summary}, we summarise the maximum achieved energy $E_{\max}$ of optimal perturbations computed for different initial energies $E_0$ and different initial time $\tau_0$, including the two cases discussed above, $\tau_0=T/2$ (left most curve) and $\tau_0=0$ (right most curve), respectively. 
We begin by discussing the low energy limit, $E_0 \rightarrow0$.
When the perturbation is applied at the end of the deceleration phase $\tau_0=3T/4$, the laminar profile still features inflection points, however, they quickly disappear during the acceleration. Linear combinations of oblique perturbations are optimal, but they can exploit the instantaneous linear instability for a much shorter time and hence experience a significantly lower growth than when perturbing at $\tau_0=T/2$. At the end of the acceleration phase ($\tau_0=T/4$), the optimal perturbation is the classical pair of streamwise vortices and shows a slightly higher growth than for $\tau_0=0$, which is attributed to the higher (peak) instantaneous Reynolds number.
\begin{figure}
    \centering
    \includegraphics[width=0.99\linewidth]{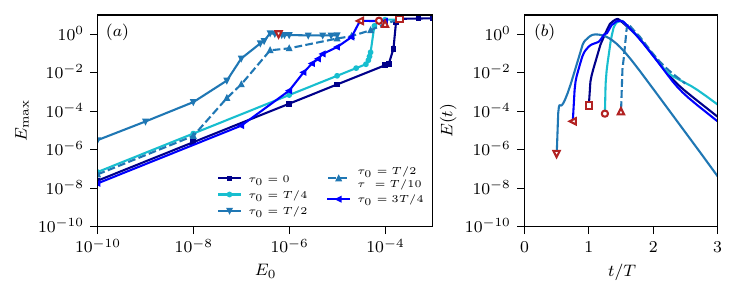}
    \caption{$(a)$ The maximum achieved energy $E_{\max}$ experienced by the optimal perturbation at different initial energies $E_0$. Solid lines refer to the different initial times of $\tau_0=0$, $\tau_0=T/4$, $\tau_0=T/2$ and $\tau_0=3T/4$ for a fixed optimisation time of $\tau=T/2$ at $(Re,Wo,A)=(2200,5.6,0.85)$, whereas the dashed line indicates the optimisation results for $\tau=T/10$
    $(b)$ The energy evolution over time of different optima that reach the saturated state with the lowest initial energy $E_0$ for the considered $\tau_0$ and $\tau$.
    }
    \label{fig: Grwoth rate summary}
\end{figure}
As $E_0$ increases, the cases $\tau_0=0$ and $\tau_0=T/4$ follow the same trend, but the growth is higher for $\tau_0=T/4$ and, accordingly, the minimal seed to trigger transition to turbulence has lower energy.
\\
The case $\tau_0=3T/4$, shows a more interesting behaviour.
For initial energies of $\mathcal{O}(E_0)=10^{-6}$, localised reflection--symmetric oblique perturbations can utilise nonlinearities to achieve an increased growth, as shown earlier for $\tau_0=T/2$.
For $\mathcal{O}(E_0)=10^{-5}$, the optimal perturbation consists of a combination of the two nonlinear optima observed for $\tau_0=T/2$ and $\tau_0=0$. This perturbation initially grows exponentially until the inflection points in the laminar profile disappear, however, the streaks generated are susceptible to further growth due to the high instantaneous Reynolds number. For $E\gtrsim 3 \cdot 10^{-5}$, the steaks are unstable and break down to a turbulent puff, exactly as for $\tau_0=0$ and $\tau_0=T/4$. Hence, turbulence is only observed in the form of puffs, in agreement with \citet{Moron2024MS}.
Once turbulence is triggered, the maximum achieved energy is independent of the initial time $\tau_0$ and initial energy $E_0$ and is reached at the same time, $t/T\approx0.45$, within a pulsation period (see figure \ref{fig: Grwoth rate summary}$b$), in agreement with the delay between bulk velocity and turbulence intensity reported by \citet{Moron2024MS}, see their figure 3$a$.

\subsection{Effect of pulsation parameters}
\label{sec: Paramterstudy}
For the parameter set considered so far, $(Re,Wo,A)=(2200,5.6,0.85)$, NLOPs experience a strong energy amplification, especially during the deceleration phase of the pulsation. However, to trigger a turbulence, a relatively large initial energy $\mathcal{O}(E_0^{crit.})=10^{-5}$ is required and turbulence is never sustained due to the long pulsation period and small minimal instantaneous Reynolds number.
In the following, we study the effect of the pulsation parameters on the transient amplification for $\tau_0=0$ and $\tau_0=T/2$, which are typical for the two distinct instability routes observed (smaller amplification and transition to turbulence at large $E_0$ ; higher amplification of symmetric oblique patterns).
\begin{figure}
    \centering
    \includegraphics[width=0.99\linewidth]{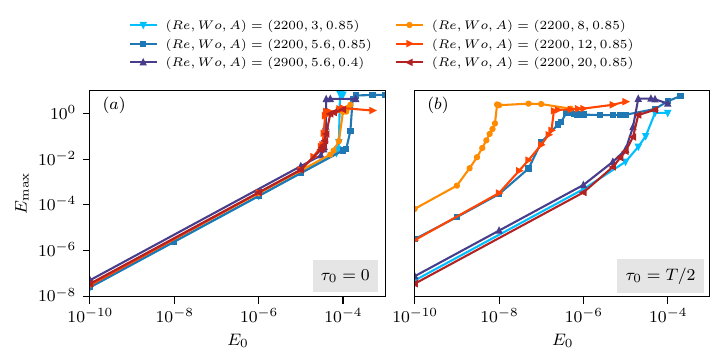}
    \caption{The maximum achieved energy $E_{\max}$ experienced by the optimal perturbation at different initial energies $E_0$ introduced in the acceleration phase $\tau_0=0$ $(a)$ and deceleration phase $\tau_0=T/2$ $(b)$. Different lines refer to different parameter combination $(Re,Wo,A)$.
    }
    \label{fig: parameterstudy}
\end{figure}
\\
Optimal perturbations introduced during the acceleration phase, $\tau_0=0$, behave qualitatively similar to steady pipe flow (see figure \ref{fig: parameterstudy}$a$). However, the specific initial energy that leads to substantial nonlinear effects and turbulence transition slightly depends on the pulsation parameters, specifically on the lengths of the period in convective time units.
\\
In the deceleration phase, $\tau_0=T/2$, optimal perturbations exploit the instantaneous linear instability of the underlying base flow. As the instantaneous linear instability depends on the pulsation parameters, the growth and dynamics of optimal perturbations are also strongly affected by $(Re,Wo,A)$, as reported by \citet{Xu2021} for the linear case.
First, we focus on the effect of the Womersley numbers. For $Wo=5.6$ and $Wo=12$ LOPs experience almost the same energy growth whereas at $Wo=8$ the growth is about two orders of magnitude higher, in agreement with \citet{Xu2021}. 
Again, NLOPs are axially localised reflection--symmetric oblique perturbations which grow by mechanisms discussed in section \S \ref{sec: NLOP Example} and further localise in axial direction with increasing initial energies.
As shown in section \S \ref{sec: effect of initial energy}, the energy growth of NLOPs for $Wo=5.6$ initially saturates at a non--turbulent state due to a combination of the long pulsation period with a small minimal instantaneous Reynolds number. 
For $Wo=8$ and $Wo=12$, the pulsation period is significantly shorter in convective time units. 
Here, viscous effects do not have enough time to dissipate the energy transferred to large wavenumber modes by nonlinear effects and a turbulent puff is triggered already at $E_0^{crit.}=9 \cdot 10^{-9}$ and $E_0^{crit.}=2 \cdot 10^{-7}$ for $Wo=8$ and $Wo=12$, respectively.
\\
At higher Womersley numbers, perturbations do not have time to exploit the instantaneous linear instability. Here, linear and nonlinear optima are as in steady pipe flow, relying solely on the Orr and lift--up mechanism. As a consequence, much higher initial energies $\mathcal{O}(E_0)=10^{-6}$ are necessary to leverage nonlinearities and the growth is significantly smaller. For $Wo=20$, $E_0\approx 2\cdot 10^{-5}$ is required to trigger transition to turbulence.
\\
For small Womersley numbers, $Wo=3$, we found the same LOPs as in steady pipe flow, in agreement with \citet{Xu2021}. By contrast, NLOPs are reflection--symmetric oblique perturbations that localise in axial and azimuthal direction and leverage short--lived inflection points in the laminar profile. Their growth is several orders of magnitude smaller than for $Wo=(5.6,8,12)$. Consequently, significantly higher $E_0$ are required to trigger transition.
For even smaller Womersley numbers, laminar profiles no longer feature inflection points and are instantaneously linearly stable throughout the period. In this quasi--steady regime, NLOPs as in steady pipe flow are expected, but their growth depends on the instantaneous Reynolds number.
Similarly, inflection points in the laminar profile disappear for sufficiently small amplitudes and the classical LOP is recovered for $A \lesssim 0.4$ \citep{Xu2021}. We performed additional optimisations at lower pulsation amplitude, while keeping the maximum Reynolds number approximately constant $(Re,Wo,A)=(2900,5.6,0.4)$ and also found LOPs as in steady pipe flow, whereas NLOPs are axially and azimuthally localised symmetric oblique perturbations.
Despite the higher mean Reynolds number, optimal perturbations at $A=0.4$ show significantly lower overall growth, which we attribute to the weakening of the inflectional profile and hence to a smaller growth by the instantaneous linear instability.

\subsection{Turbulence dynamics}
\label{sec: Turbulent dynamics}
We performed DNS for a given parameter set to illustrate the dynamics of turbulence after transition. As initial condition we used the minimal seed, i.e. the optimal perturbation that triggers transition at lowest $E_0$. 
\begin{figure}
    \centering
    \includegraphics[width=0.99\linewidth]{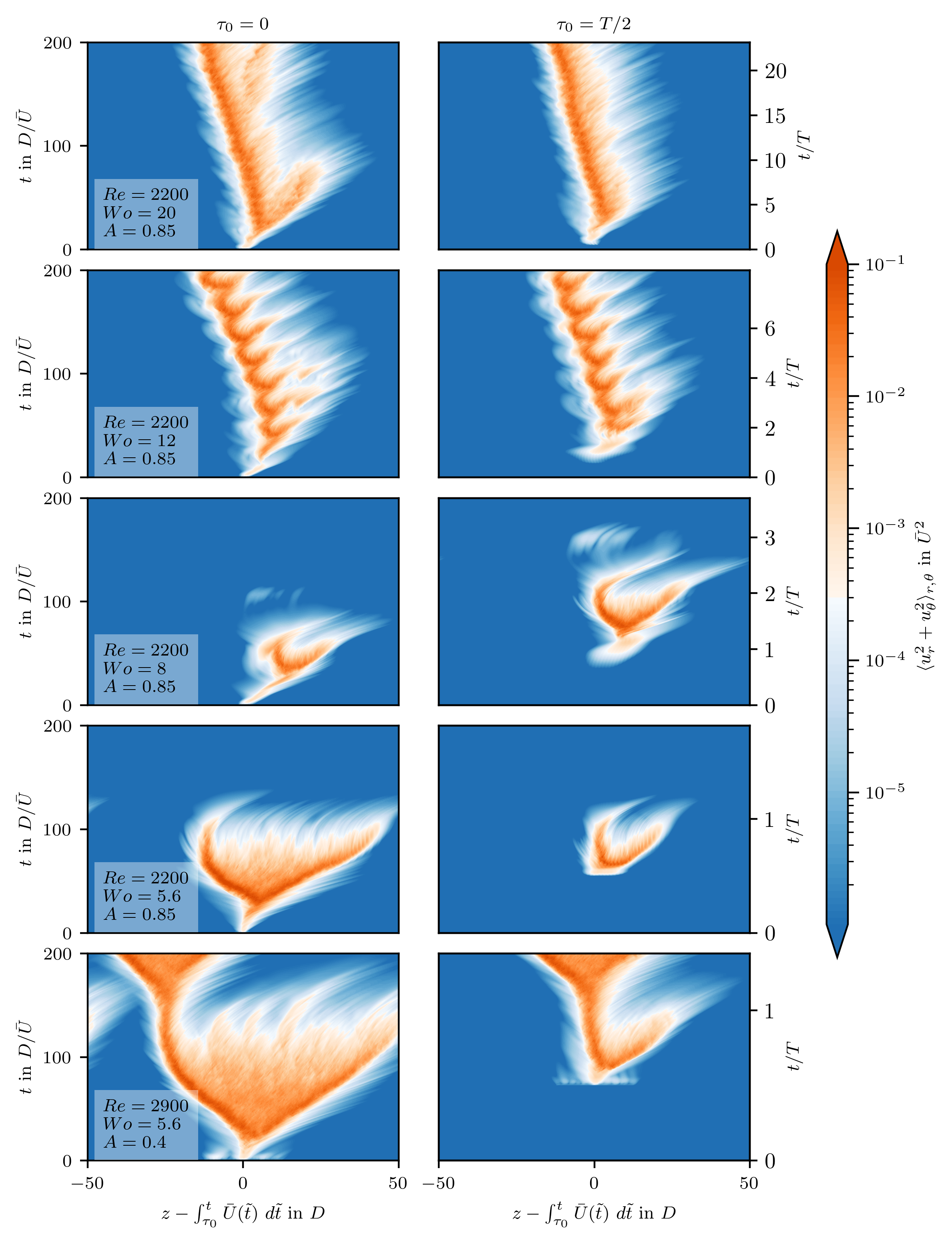}
    \caption{Space--time diagrams of the cross--sectionally averaged turbulent cross--section kinetic energy of minimal seeds in a frame of reference co--moving with the bulk velocity. The left and right column depict the evolution of optimal perturbations introduced in the acceleration phase $\tau_0=0$ and deceleration phase $\tau_0=T/2$, respectively. To delay the interaction of puffs/slugs due to the periodic boundary conditions, the pipe length was set to $L_z=200$ with a resolution of $(N_r,M,K)=(96,40,1200)$.
    }
    \label{fig: space time diagramms}
\end{figure}
In figure \ref{fig: space time diagramms}, we show space--time diagrams of the cross--sectionally integrated cross--section kinetic energy at different parameter combinations in a frame of reference co--moving with the bulk velocity. 
Once a puff has been triggered, its dynamics is governed by the instantaneous Reynolds number $Re(t)$ of the pulsation \citep{Xu2018}. 
For $(Re,Wo,A)=(2200,12,0.85)$ and $(2200,20,0.85)$, the phase of instantaneously low $Re(t)$ is short. In this high frequency regime, the puff dynamics is modulated by the pulsation but puffs have neither time to decay nor to split (although splitting would be expected here for sufficiently long time). Although the energetically more efficient transition scenario depends on $\tau_0$, once triggered, the turbulence dynamics are independent of $\tau_0$ and $E_0$.
\\
As $Wo$ is decreased, the length of the period increases, and for $Wo=8$, puffs first expand and then decay completely. When the perturbation is applied during the deceleration phase $\tau_0=T/2$, puffs emerge during the phase of high instantaneous Reynolds number and decay during the subsequent low Reynolds number phase. Compared to $\tau_0=0$, transition is not only energetically more favourable, but puffs reach slightly higher turbulent kinetic energies and spread fuhrer in axial direction. 
A similar trigger--decay sequence is observed for $Wo=5.6$, however, the period is much longer and for both $\tau_0=0$ and $\tau_0=T/2$, puffs decay during the first low Reynolds number phase. Puffs triggered at $\tau_0=0$ evolve into expanding slugs during the long phase of large instantaneous Reynolds numbers and reach significantly higher amplitudes than for $\tau=T/2$, where puffs decay quickly.
When the pulsation amplitude is decreased, while keeping approximately the same $Re_{\max}$, optimal perturbation trigger strongly expanding slugs during large $Re(t)$.  
However, because of the higher $Re$ and lower $A$, the time of low instantaneous $Re(t)$ is significantly shorter. Puffs survive the low Reynolds number phase and develop into a slug again during the next acceleration phase. From this time on, dynamics are independent of $\tau_0$ and $E_0$.

\section{Conclusion}
\label{sec: Discussion}
Our nonlinear stability analysis demonstrates that pulsatile pipe flow at moderate pulsation amplitudes $0.4\lesssim A \lesssim 1$ is subject to two distinct instability routes.
During the deceleration phase and at intermediate Womersley numbers, the flow is susceptible to the large growth of oblique perturbations, in agreement with experiments by \citet{Xu2020} and linear computations and DNS by \citet{Moron2022}. In contrast, during the acceleration phase or at low or high Womersley numbers, nonlinear optimal perturbations are similar to those of steady pipe flow \citep{Pringle2010, Kerswell2014}.
\\
We show that at intermediate $Wo$, instability is energetically much more favourable in the deceleration phase. The optimal perturbation is axially localised and consists of symmetric pairs of oblique (helical) modes. These NLOPs can grow by up to nine order of magnitude in energy despite the moderate Reynolds numbers considered ($Re=2200, \ Re_{\max}=4070$). They exploit nonlinearities to form strong axial streaks due to the lift--up mechanism, after initially leveraging the Orr mechanism and the instantaneous linear instability of the underlying base flow. This perturbation generates, regular flow patterns that subsequently decay or break down to turbulence while being advected downstream. 
Their huge nonlinear growth reported here explains why instability was cyclically observed in the experiments of \citet{Xu2020}. However, they reported flow patterns with a preferred helical direction, whereas here we found that perfectly helical perturbations are nonlinearly suboptimal. Specifically, they are unable to leverage nonlinearities. Our findings are consistent with the DNS of \citet{Feldmann2020}, who observed helical patterns only when their perturbation had a preferred direction, and suggests that the pipe bend used in \citet{Xu2020} as perturbation may have broken the azimuthal reflection symmetry of the system. 
\\
During the acceleration phase or at small or large $Wo$, optimal perturbations are similar to those observed in steady pipe flow and require significantly higher initial energies than in the deceleration phase (up to four orders of magnitude more) to exploit nonlinear effects. 
These perturbations trigger a turbulent puff that then grows into a slug and eventually shrinks and decays depending on the flow parameters.
\\
All in all it can be concluded that pulsatile pipe flow at moderate pulsation amplitudes and Reynolds numbers is extremely prone to transition. 
At low disturbance levels (and intermediate $Wo$), turbulence is triggered during the deceleration phase due to the inflectional laminar profile. 
If the disturbance level is high, the classical transition scenario as in steady pipe flow prevails and turbulence can also emerge during the acceleration phase. 
Similarly, the classical scenario is recovered in the quasi--steady and high frequency limits.
Once triggered, turbulence survives if $Wo$ or $Re_{\min}$ is high enough \citep{Xu2017,Xu2018}. Otherwise the flow relaminarizes and turbulence is retriggered every cycle.
\\
We note that the parameter ranges relevant to the large arteries in the human cardiovascular system fall within the regime of intermediate Womersley numbers and high pulsation amplitudes \citep{Quarteroni2019}.
The blood vessels are vastly more complex than the rigid, smooth and straight pipe investigated here, naturally giving rise to disturbances. We thus suggest that strong nonlinear patterns and turbulence may be more ubiquitous in the cardiovascular system than usually assumed.

\section*{Acknowledgement}\label{Ack}
\noindent
This work was funded by the Deutsche Forschungsgemeinschaft (DFG) in the framework of the research unit FOR 2688 'Instabilities, Bifurcations and Migration in Pulsatile Flows' under grant number 349558021, which is gratefully acknowledged. Further, the authors acknowledge the fruitful discussions with Dr. Daniel  Mor\'on.
\section*{Deceleration of Interest} 
The authors report no conflict of interest.
\section*{Author ORCIDs} 
\orcidlink{0009-0005-6141-3145} Patrick Keuchel \href{https://orcid.org/0009-0005-6141-3145}{\textcolor{blue}{https://orcid.org/0009-0005-6141-3145}}; \\
\orcidlink{0000-0001-5988-1090} Marc Avila \href{https://orcid.org/0000-0001-5988-1090}{\textcolor{blue}{https://orcid.org/0000-0001-5988-1090}}.
\bibliographystyle{jfm}
\bibliography{jfm.bib}
\appendix
\section{Code validation}
\label{app: Validation pulsatile}
We validated our implementation of the adjoint looping procedure, by computing optimal perturbations in the nonlinear regime of steady pipe flow \citep{Pringle2010, Pringle2012} and in the linear regime of pulsatile pipe flow \citep{Xu2021}.
\subsection{Optimal perturbations in steady pipe flow}
First, nonlinear optima were computed in short $L_z=\pi$ long pipes at $Re = 1750$ for Hagen--Poiseuille flow \citep{Pringle2010}. We set the optimisation time to $\tau=80$ and initialize the optimization procedure with a turbulent field, obtained for the same Reynolds number and domain, rescaled to the initial energy of $E_0$. 
\begin{figure}
\centerline{\includegraphics[width=.99 \textwidth]{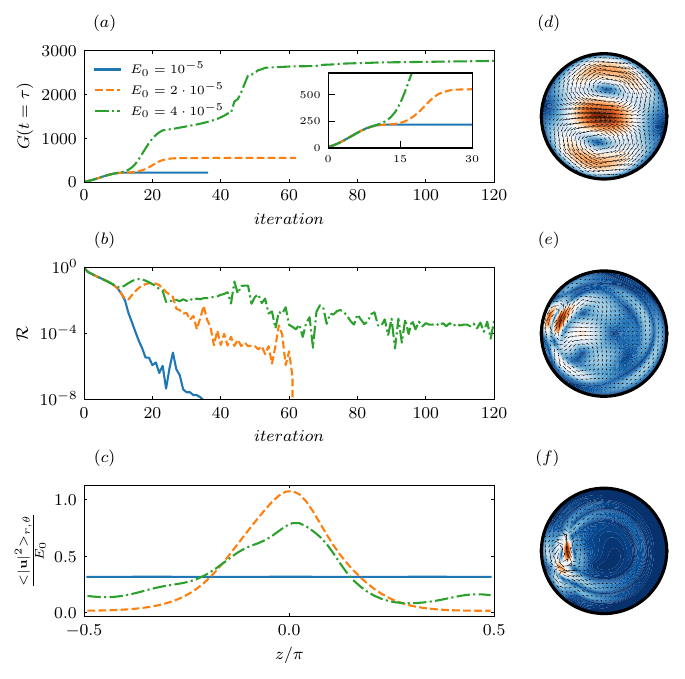}}
\caption{The evolution of the growth rate $G$ at the optimisation time $t=\tau$ over the adjoint iteration $(a)$, together with the relative incremental growth rate change $\mathcal{R}=|(G^{s+1}-G^{s})/G^{s+1}|$ $(b)$ and the relative axial distribution of cross--sectionally integrated energy $(c)$ for initial energies $E_0 =1 \cdot 10^{-5}$ (blue/--$\!$--), $E_0 =2 \cdot 10^{-5}$(orange/$-$$-$$-$) and $E_0 =4 \cdot 10^{-5}$ (green/$- \cdot - \cdot -$). $(d)-(f)$ Contour of the cross--sectional velocity $\sqrt{u_r^2+u_{\theta}^2}$ for the optimal initial perturbation with an energy of $E_0=10^{-5}$, $E_0=2 \cdot 10^{-5}$ and $E_0=4\cdot 10^{-5}$ respectively. The axial position corresponds to the position of maximum cross--sectionally integrated energy. All simulations were performed at $Re=1750$, $L_z=\pi$ with a spatial resolution of $(N_r,M,K)=(38,16,21)$ and a time step of $\Delta t=0.01$.}
\label{Fig: NLOPS LzPi Re1750 Initial}
\end{figure}
In figure \ref{Fig: NLOPS LzPi Re1750 Initial}$a$, we plot the evolution of the energy gain at the optimisation time, $G(t=\tau)$, over the adjoint looping iterations for three initial energies. 
For the lowest initial energy of ($E_0=10^{-5}$), the adjoint looping procedure quickly improved the initial perturbation and after $s\approx 10$ iterations, the initial perturbation smoothly approached an optimum. The procedure was terminated after the relative improvement of the growth fell below a threshold, here $\mathcal{R}=|(G^{s+1}-G^{s})/G^{s+1}|<10^{-8}$ (see figure \ref{Fig: NLOPS LzPi Re1750 Initial}$b$). This is a very conservative limit and does not reflect the accuracy of the determined optimal growth, i.e. discretisation errors are larger. However, we have observed that the optimisation can stagnate at relatively low levels, e.g. $\mathcal{R}\approx 10^{-5}$ before a further substantial improvement occurs, especially for longer pipes. A similar behaviour was observed by \citet{Pringle2012}. In figure \ref{Fig: NLOPS LzPi Re1750 Initial}$c$, we show the cross--sectionally integrated energy of the initial perturbation as a function of $z$. The optimal perturbation corresponds to the linear optimum, consisting of a pair of streamwise independent vortices ($m=1,\ k=0$), shown in figure \ref{Fig: NLOPS LzPi Re1750 Initial}$d$.
Integrating this optimal perturbation in time, leads to the well known formation of streaks due to lift--up effect (not shown here). These streaks grow in strength and the perturbation energy reaches a maximum amplification of $G_{\max} \approx 219$ at $t=85.05$ (see figure \ref{Fig: NLOPS LzPi Re1750 Evolution}$a$), in line with previous studies \citep{Meseguer2003, Pringle2010}. Due to the absence of three--dimensional fluctuations, the streaks remain stable and ultimately, viscous effects lead to an exponential decay of the perturbation.
\begin{figure}
    \centering
    \includegraphics[width=.99 \textwidth]{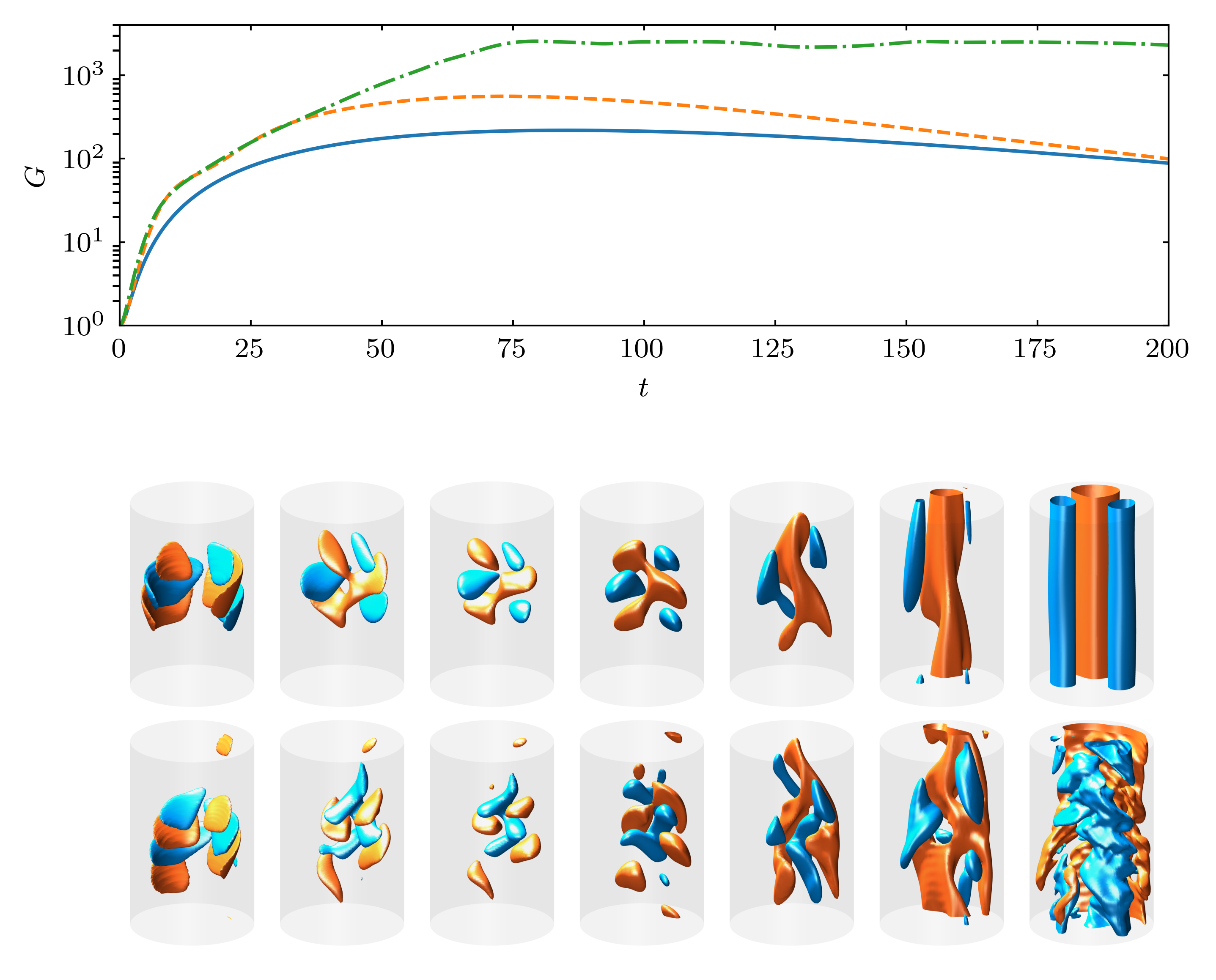}
    \caption{$(a)$ Evolution of the growth rate over time for the initial energies $E_0 =1 \cdot 10^{-5}$ (blue/--$\!$--), $E_0 =2 \cdot 10^{-5}$ (orange/$-$$-$$-$) and $E_0 =4 \cdot 10^{-5}$ (green/$- \cdot - \cdot -$). $(b)$ Evolution of the NLOP with initial energy of $E_0=2 \cdot 10^{-5}$ at time $t=0$, $t=1.6$, $t=4$, $t=10$, $t=20$, $t=40$ and $t=80$ (from left to right). The mean flow is from bottom to top and isocontours depict $+30\%$ (orange) and $-30\%$ (blue) of the maximum axial perturbation velocity. $(c)$ same as $(b)$ for $E_0=4 \cdot 10^{-5}$}
\label{Fig: NLOPS LzPi Re1750 Evolution}
\end{figure} 
\\
For an increased initial energy ($E_0 =2 \cdot 10^{-5}$), the optimisation procedure identified an optimum after about $s\approx 30$ iterations, which experienced a  significantly higher growth. In figure \ref{Fig: NLOPS LzPi Re1750 Initial}$c$ and \ref{Fig: NLOPS LzPi Re1750 Initial}$e$ we show that this perturbation localizes in all directions.
As shown in figure \ref{Fig: NLOPS LzPi Re1750 Evolution}, this nonlinear optimal perturbation (NLOP) exploits the Orr mechanism to achieve an initial energy boost. Once the strong axial velocity layers align with the shear, oblique waves dominate the flow. The perturbation further gains energy in this second (oblique wave) phase and during its nonlinear evolution, energy is transferred to the streamwise independent modes. This initiates the third phase, in which the flow becomes increasingly two--dimensional and the energy grows due to the lift--up effect.
Sequentially utilising these various growth mechanisms, allows this NLOP to reach a significantly higher maximum growth of $G_{\max}=560.6$ at $t=73.4$, but ultimately the perturbation decays. 
\\
By further increasing the initial energy ($E_0 =4 \cdot 10^{-5}$), an optimum with even higher growth was identified, however, the optimisation procedure did not converge in this case (we continued the optimisation procedure for $s=350$ iterations). The perturbation computed in the last iteration is qualitatively similar to the the NLOP at $E_0 =2 \cdot 10^{-5}$ (see figure \ref{Fig: NLOPS LzPi Re1750 Initial}$c$ and \ref{Fig: NLOPS LzPi Re1750 Initial}$f$). In figure \ref{Fig: NLOPS LzPi Re1750 Evolution}, we plot the temporal evolution of energy (top panel) and isocontours of the axial velocity (bottom panel). This perturbation causes significantly stronger streaks, which become unstable and turbulence is triggered. It is above the minimal seed \citep{Pringle2010}, where many perturbations of energy $E_0$ are capable of triggering turbulence. Since the kinetic energy of the turbulent flow changes with time, the precise value of $G$ obtained depends sensitively on the initial condition and no convergence occurs \citep{Pringle2012}.
\\
The computed NLOPs show larger growth, than those reported by \cite{Pringle2010}, however, their structure and evolution are qualitatively in good agreement. We suspect that the difference in the growth originates from the localization in axial direction, which was not found in \citet{Pringle2010} and \cite{Pringle2012} for short domains.
 \\
An axially localized nonlinear optimum was previously reported in \citet{Pringle2012} for longer pipes ($L_z=10$) at $\Rey=2400$ and $\tilde{E}_0=E_0/E_{lam}=7.077\cdot 10^{-6}$. We performed an optimization at the same parameters and also find a localized perturbation (see figure \ref{fig: Validation Lz10}). 
\begin{figure}
    \centering
    \includegraphics[width=0.99\linewidth]{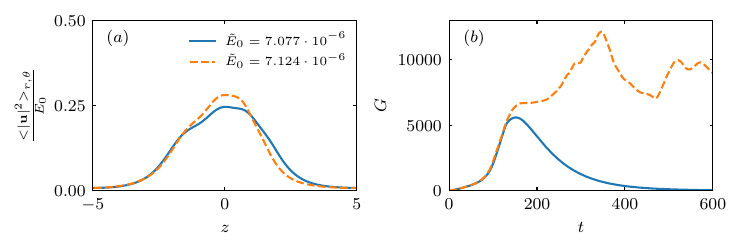}
    \caption{Axial distribution of cross--sectionally integrated energy, normalized with $E_0$, $(a)$ and growth over time $(b)$ for the initial energies $\tilde{E}_0=E_0/E_{\text{lam}}=7.077\cdot 10^{-6}$ (blue/--$\!$--) and $\tilde{E}_0=E_0/E_{\text{lam}}=7.124\cdot 10^{-6}$ (orange/$-$$-$$-$) at $Re=2400$, $L_z=10$.}
    \label{fig: Validation Lz10}
\end{figure}
Slightly increasing the initial energy to $\tilde{E}_0=7.124\cdot 10^{-6}$ triggers turbulence transition. Thus the minimal seed for this setup lies in--between $7.077\cdot 10^{-6} \leq \tilde{E}_0 \leq 7.124\cdot 10^{-6}$ (see figure \ref{fig: Validation Lz10}), which is in agreement with the results reported by \citet{Pringle2012}.

\subsection{Linear optimal perturbations in pulsatile pipe flow}
\begin{figure}
    \centering
    \includegraphics[width=0.99 \textwidth]{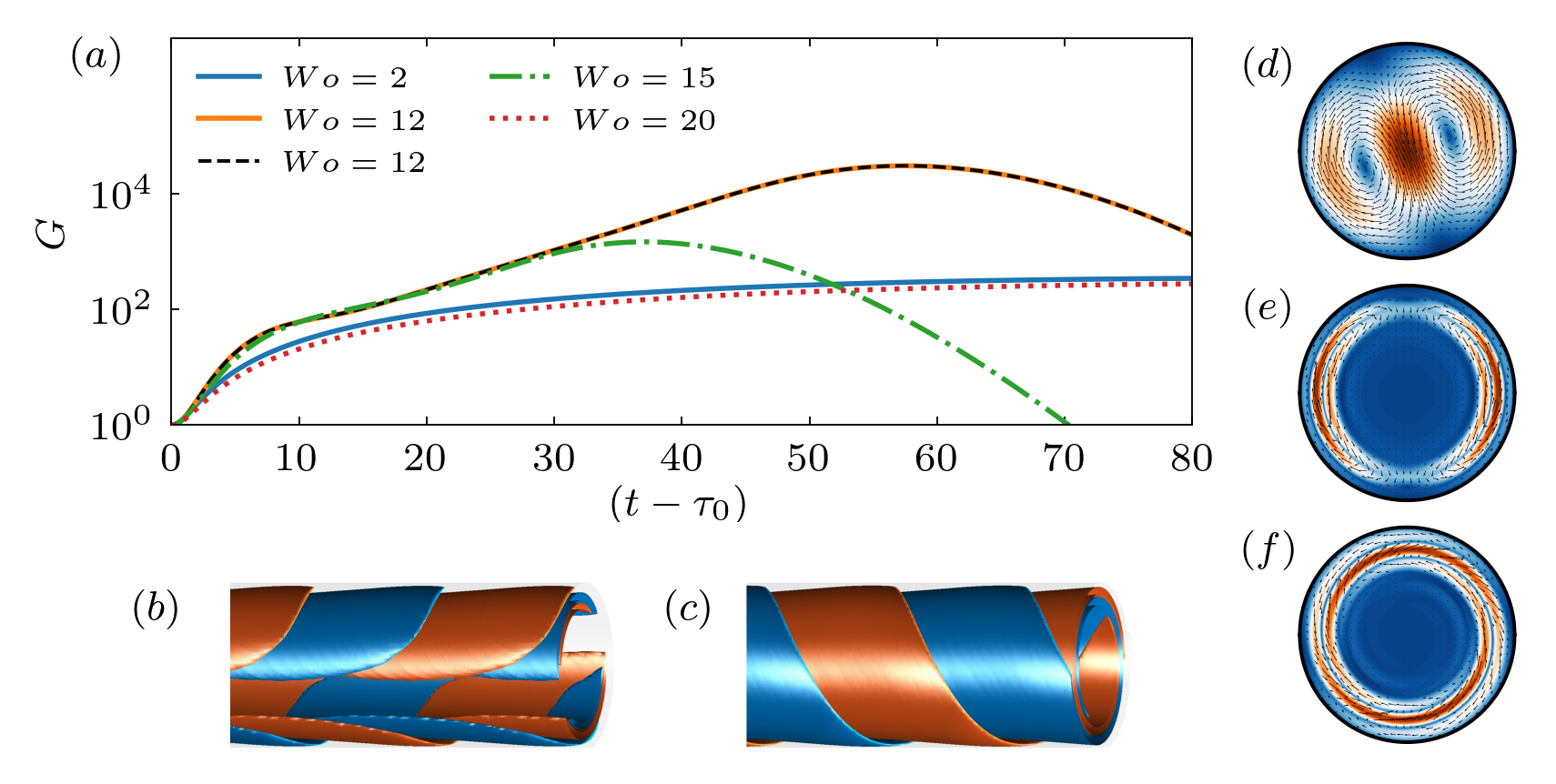}
    \caption{$(a)$ Growth over time for $Wo=2$ (blue/---$\!$---), $Wo=12$ (orange and black/$---$), $Wo=15$ (green/$-\cdot-\cdot -$) and  $Wo=20$ (red/$\cdots \cdots$) at $(Re,A)=(2000,1)$. For $Wo=12$ (orange and black/$---$), we show the growth of a symmetric pair of oblique waves (black) and a helical wave perturbation (orange). $(b),(c)$ Isocontours of $\pm30\%$ of the maximum axial velocity of a symmetric pair of oblique waves and a helical wave perturbation respectively. $(d)-(f)$
    Contours of the cross--sectional velocity $\sqrt{u_r^2+u_{\theta}^2}$ and quivers of the cross--sectional components $u_r, \ u_\theta$ for the optimal streamwise vortices, symmetric pairs of oblique waves and a helical wave respectively. All these optimisations were carried out with a constant time step of $\Delta t = 0.01$ and a spatial resolution of $(N_r,M,K)=(38,16,64)$.}
    \label{fig: validation helical}
\end{figure}

As shown by \citet{Xu2021}, a new linear optimal emerges for pulsation frequencies $4\lesssim  Wo \lesssim 
18$ and pulsation amplitudes $A\gtrsim 0.4$. This optimal perturbation has a helical shape and utilizes the Orr mechanism and the instantaneous linear instability of the laminar profile \citep{Moron2022} to outgrow the classical linear optimal perturbation (pair of streamwise independent vortices) by many orders of magnitude \citep{Xu2021}. 
To validate our code for pulsatile base flows, we computed optimal perturbations in the linear regime ($E_0=1\cdot 10^{-20}$) at $Re=2000$, $A=1$, $L_z=10$ for Womersley numbers $Wo\in[2,12,15,20]$. Based on the results in \citet{Xu2021}, we introduce the perturbation at time $\tau_0\approx T/2$ and aim to maximise the growth at the final time $t_f=\tau_0+\tau$. Here, we set the optimisation times of the adjoint looping procedure to $\tau\in[0.02T,0.7T,0.7T,3.5T]$ for $Wo\in[2,12,15,20]$ respectively, where $T=\pi Re/(2Wo^2)$ is the dimensionless period.
\\
In figure \ref{fig: validation helical}, we plot the growth evolution of the computed optimal perturbations, together with isocontours of the axial velocity and contours of the cross--sectional velocity.
For a low Womersley number of $Wo=2$, we obtained the classical optimal consisting of a pair of streamwise vortices (see figure \ref{fig: validation helical}$d$). With a pulsation period much higher than the convective time scale, this perturbation grows on top of the quasi steady profile and reaches a maximum amplification at $t\approx 90$ $(0.029T)$.
\\
Increasing the Womersley numbers to $Wo=12$ and $Wo=15$, led to qualitatively different optima, which experienced an exponential growth phase and ultimately reached a much higher energy gain.
The optimum at $Wo=12$ is a symmetric pair of oblqiue waves with an equal contribution of the azimuthal modes $m=1$ and $m=-1$ (see figure \ref{fig: validation helical}$b$ and \ref{fig: validation helical}$e$). However, depending on the initial guess, we also found a helical wave with azimuthal wave number of $m=1$ or $m=-1$ (see figure \ref{fig: validation helical}$c$ and \ref{fig: validation helical}$f$) as in \citet{Xu2021}, or any linear combination to be equally optimal. 
The energy of these optimal perturbations is concentrated in the axial mode $k=8$, corresponding to a structure that repeats eight times in the axial (periodic) direction, i.e. with an axial length of $10/8$ radii. 
During the first stages, this perturbation experiences a strong initial energy boost via the Orr mechanism. In this process, the axial velocity layers tilt until they align with the shear.
Once the velocity layers align with the shear, the perturbation experiences an exponential growth (see figure \ref{fig: validation helical}$a$). The exponential growth is attributed to the presence of inflection points of the laminar profile during the deceleration phase and its instantaneous linear instability \citep{Moron2022}. In the linear regime, the entire energy remains in the initially excited modes during the perturbation's evolution and hence, the structure of the flow remains qualitatively similar.  
\\
Qualitatively, the same optima were found for $Wo=15$. The growth of these optimal perturbation is lower compared to $Wo=12$, due to the shorter period \citep{Moron2022}. 
Further increasing the Womersley number to $Wo=20$ leads to much shorter period length and the optimisation recovers the classical optimal perturbation. 
All results are in excellent agreement with the results in \citet{Xu2021}. 
\end{document}